\documentclass[12pt,a4paper,reprint,aip,pof,onecolumn]{revtex4-1}
\usepackage{amsbsy,graphicx,textcomp,tikz,todonotes}

\usepackage{graphicx, epsfig, color, amsmath}
\usepackage[left=1.in,right=1.in,top=1in,bottom=1in]{geometry}
\usepackage{subfigure}
\usepackage{natbib}
\begin{document}
\title{Sample dispersion in isotachophoresis with Poiseuille counterflow}

\author{Somnath Bhattacharyya}
%\email[]{somnnath@maths.iitkgp.ernet.in}
\affiliation{Department of Mathematics, Indian Institute of Technology, 721302 Kharagpur, India}

\author{Partha P. Gopmandal}
%\email[]{parthamaths@gmail.com}
\affiliation{Department of Mathematics, Indian Institute of Technology, 721302 Kharagpur, India}

\author{Tobias Baier}
\email[Corresponding author: ]{baier@csi.tu-darmstadt.de}
\affiliation{Center of Smart Interfaces, Petersenstr. 32, 64287 Darmstadt, Germany}

\author{Steffen Hardt}
%\email[]{hardt@csi.tu-darmstadt.de}
\affiliation{Center of Smart Interfaces, Petersenstr. 32, 64287 Darmstadt, Germany}

\date{\today}

\begin{abstract}
A particular mode of isotachophoresis (ITP) employs a pressure-driven flow opposite to the sample electromigration direction in order to anchor a sample zone at a specific position along a channel or capillary. We investigate this situation using a two-dimensional finite-volume model based on the Nernst-Planck equation. The imposed Poiseuille flow profile leads to a significant dispersion of the sample zone. This effect is detrimental for the resolution in analytical applications of ITP. We investigate the impact of convective dispersion, characterized by the area-averaged width of a sample zone, for various values of the sample P\'{e}clet-number, as well as the relative mobilities of the sample and the adjacent electrolytes. A one-dimensional model for the area-averaged concentrations based on a Taylor-Aris-type effective axial diffusivity is shown to yield good agreement with the finite-volume calculations. This justifies the use of such simple models and opens the door for the rapid simulation of ITP protocols with Poiseuille counterflow.
\end{abstract}

\maketitle

\renewcommand {\baselinestretch} {1.25} \normalsize
\fontsize{12pt}{15pt}\selectfont

\section{Introduction\label{sec:Intro}}
Isotachophoresis (ITP) is a special mode of electrophoretic transport that has already been described by Kohlrausch \cite{1}. It has found widespread applications in the analytical sciences (for an overview, see reference\cite{2}) and is mainly used as a technique for the preconcentration and separation of samples. ITP relies on consecutively stacking a high mobility leading electrolyte (LE) and a low mobility trailing electrolyte (TE) in a capillary or channel as sketched in figure \ref{fig:Sketch}. Upon application of an electric field the ions arrange in the order of their electrophoretic mobility, forming a sharp transition zone between them that migrates along the capillary according to the electrophoretic velocity of the ions. Accordingly, there is a corresponding change in electric field across this transition zone, such that the high and low mobility species move at the same speed. A high mobility ion diffusing backwards across the transition zone into the low mobility region will experience this higher electric field and is thus transported back to its own region and vice versa for a low mobility ion diffusing into the region of high mobility ions. It is this balance between diffusion and electrophoretic migration that determines the width of the transition zone. Depending on whether the stacking occurs for the anions or the cations, anionic or cationic ITP is observed. Sample ions with an electrophoretic mobility in between the LE and TE ion mobilities are sandwiched between the two electrolytes and can form a zone of only a few micrometers width, depending on the total amount of sample.

Very early in the development of electrophoretic separation techniques researchers applied a counterflow opposing the electrophoretic sample migration to increase the time span available for the separation of analytes or to reduce the length of the separation passage \cite{3} . While this principle was first applied to capillary electrophoresis, Everaerts et al. \cite{4} introduced a counterflow to balance the sample migration in ITP experiments. Originally, pressure-driven counterflow was applied, but later electroosmotic counterflow was considered too \cite{5}. In the past decades, counterflow-balanced ITP has found widespread applications, for example in the context of sample preconcentration for improving the sensitivity of capillary electrophoresis {\cite{6}$^-$\cite{8}} or to control the elution of analytes \cite{9}.

Despite the importance of the method, it is rather poorly understood how the counterflow affects the analyte distribution in a sample focused by ITP. Urb\'anek et al.\cite{10} have noted in their experiments that pressure-driven counterflow increases sample dispersion. A few efforts have been made to compute the effects of counterflow on ITP. The one-dimensional models that have been developed \cite{11}$^,$\cite{12} may be suitable for flow profiles resembling a plug flow (such as electroosmotic flow in capillaries that are much wider than the Debye layer thickness), but do certainly not account for the complex convection-diffusion-electromigration phenomena occurring when a Poiseuille counterflow is applied. More successful one-dimensional models considering convective dispersion in ITP were developed with the aim to describe dispersion in a setting with a mismatch in electroosmotic flow \cite{Saville_1990, Ghosal_2003, 21,22}.

In this article, we present the first computational fluid dynamics (CFD) study of sample dispersion occurring when ITP is balanced by a Poiseuille counterflow. For this purpose we have numerically solved the coupled  Nernst-Planck and charge conservation equations using a finite-volume method. We analyze the most important dependencies of the width of the area-averaged sample distribution on the dimensionless parameters governing the problem. We show that in ITP with short sample zones and at large enough (but not unrealistic) P\'{e}clet numbers the flow considerably broadens the sample distribution compared to pure ITP without counterflow. Furthermore, we show that a one-dimensional Taylor dispersion model reproduces the data obtained with CFD reasonably well.

\section{Governing equations\label{sec:GovEq}}

The migration of charged species in an electrolyte under an external electric field is governed by convection, diffusion and electromigration. The mass conservation of the ionic species leads to the Nernst-Planck equation,
\begin{equation}
\frac{\partial C_i}{\partial t} +\nabla\cdot\left[\left(\textbf{V}+z_i\mu_i\textbf{E}\right)C_i-D_i \nabla C_i\right]=0.
\label{eq:NernstPlanck}
\end{equation}
Here $C_i$, $D_i$, $\mu_i$, $z_i$ denote, respectively, the concentration, diffusion coefficient,  mobility and valency of the $i^{th}$ ionic species. Our model system comprises four ionic species: ions in the TE, sample and LE having a valency of the same sign and a common counter-ion with valency of opposite sign. The subscript $i$ denoting these species is chosen as $t$, $s$, $l$, and $0$, respectively. The mobility and diffusivity are related via the Nernst-Einstein relation $\mu_i=D_i F/RT$, where $F$ is the Faraday constant, $R$ the gas constant and $T$ the absolute temperature. $\textbf{V}$ is the flow velocity field and $\textbf{E}=-\nabla \phi$ the electric field. The electric current density and charge density are defined as $\textbf{j}=F\sum_i{z_iN_i}$ and $\rho_e=F\sum_i{z_i C_i}$, where $\textbf{N}_i=-D_i \nabla C_i+C_i( \textbf{V}+z_i\mu_i\textbf{E} )$ denotes net flux of the  $i^{th}$ ionic species with $z_i=1$ for $i=t,s$ or $l$ and $z_0=-1$.

The charge conservation equation is
\begin{equation}
\frac{\partial \rho_e}{\partial t}+ \nabla\cdot\textbf{j}=0.
\label{eq:ChargeConsTime}
\end{equation}
Under electro-neutrality\cite{13}, i.e. $\rho_e=0$, the time-dependent and convective terms in (\ref{eq:ChargeConsTime}) vanish. By multiplying equation (\ref{eq:NernstPlanck}) by $Fz_i$ and taking the sum over all species the reduced charge conservation equation can then be written as
\begin{equation}\label{eq:ChargeCons}
\nabla\cdot(\nu \textbf{E})=\nabla\cdot(F\sum_i{D_iz_i\nabla C_i}),
\end{equation}
where the ionic conductivity is  $\nu=(F\sum_i{\mu_i z_i^2 C_i})$. The right-hand side term in (\ref{eq:ChargeCons}) is known as the diffusion current and its contribution is insignificant at all locations, except for the transition zones where large concentration gradients occur. For microchannels with charged walls, electroneutrality is violated within a region of the order of the Debye screening length, $\lambda_D$, much smaller than the channel dimensions. In such cases equation (\ref{eq:ChargeCons}) is retained in the channel bulk, while the effect of the charge layer is incorporated via an effective wall-velocity boundary condition taking into account the electroosmotic flow\cite{Schoenfeld_2009}. Since here we focus on sample dispersion by Poiseuille flow we assume uncharged walls.

\begin{figure}
\includegraphics[width=11 cm]{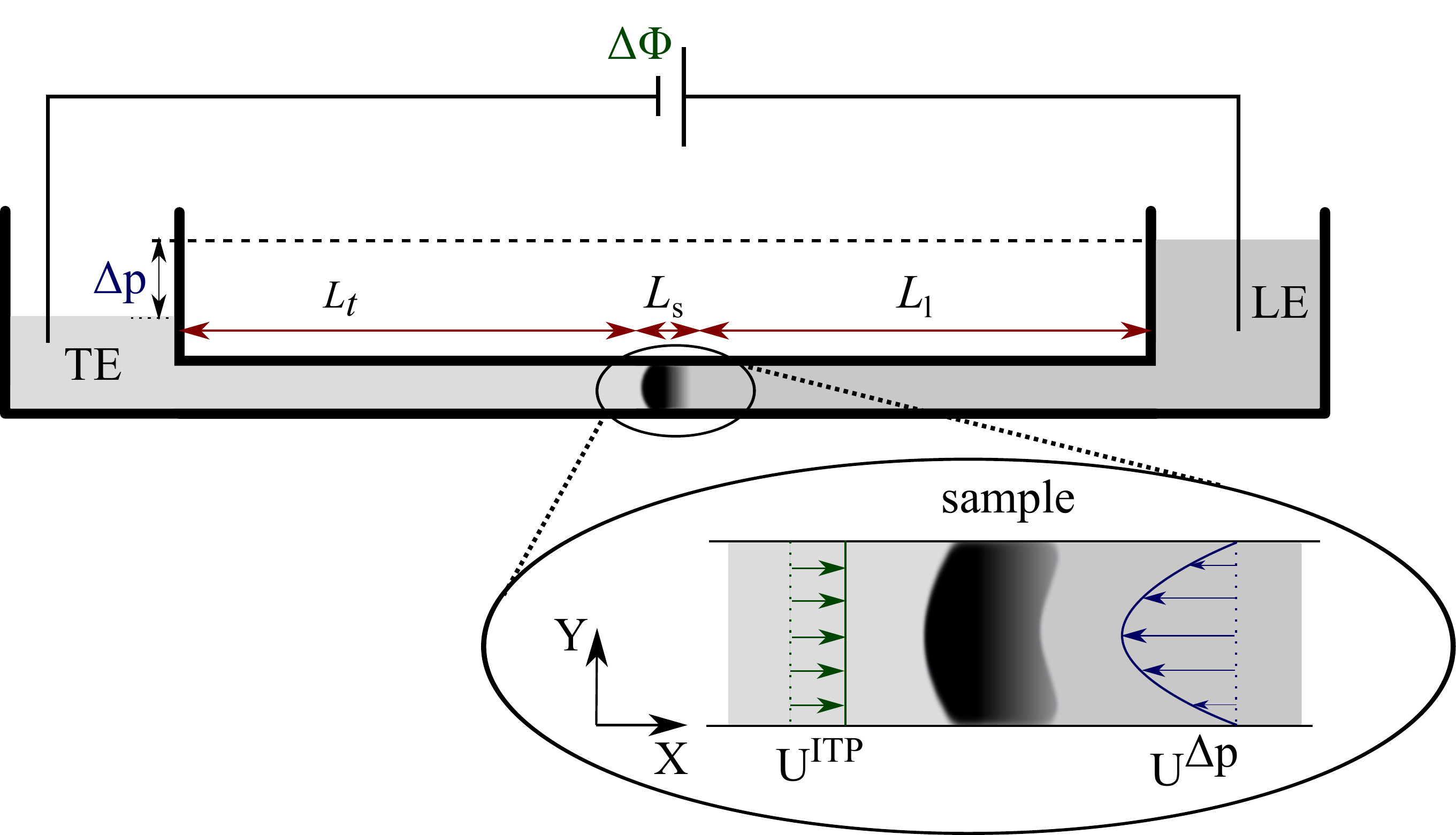}
\caption{\label{fig:Sketch} Sketch of the situation considered. A single sample zone is sandwiched between a leading (LE) and trailing (TE) electrolyte. The sample ions migrate from left to right at velocity $U^{ITP}$ in an applied electric field. A pressure driven flow from right to left is applied that exactly counters the electromigration, such that the sample plug is stationary. The channel length, $L=L_{t}+L_{s}+L_{l}$, is the sum of lengths occupied by the leading and trailing electrolytes as well as the sample. We will consider $L_s \ll L_t,\, L_l$.}
\end{figure}

In this article we consider a two-dimensional situation, i.e. the isotachophoretic transport and the Poiseuille flow occur between parallel plates. Translated to realistic microfluidic setups this means that very shallow channels with a width much larger than their depth are assumed. We impose a Poiseuille counterflow with an average speed equal and opposite to ITP migration, as
\begin{equation}
u(Y)=-6 U^{ITP}\frac{Y}{H}\left(1 -\frac{Y}{H}\right),
\end{equation}
where $H$ is the depth of the channel and $U^{ITP}$ is the velocity of ITP transport for the case that no counterflow is applied. The situation is depicted in figure \ref{fig:Sketch}. Thus, on average the electromigration of sample ions is just compensated by convection, meaning that the average speed of the ITP transition zone (the ``interface" between LE and TE) will be zero.

%%%%%%%%%%%%%%%%%%%%%%%%%%%%%%%%%%
%%%%%%%%%%%%%%%%%%%%%%%%%%%%%%%%%%

We non-dimensionalize the $X$- and $Y$-coordinates by the depth $H$ of the channel, i.e. set $x=X/H$ and $y=Y/H$. Further, all concentrations are non-dimensionalized by the bulk LE concentration, $C^\infty_{l}$, and the electric potential by $\phi_0=RT/F$. We choose the isotachophoretic velocity $ U^{ITP}$ as velocity scale. Time is scaled by $\tau=H/U^{ITP}$, i.e. $\hat{t}=t/\tau$. The Nernst-Planck equation (\ref{eq:NernstPlanck}) can then be written in non-dimensionalized form as
\begin{equation}
\frac{\partial c_i}{\partial \hat{t}} + u \frac{\partial c_i}{\partial x}  - z_i\frac{D_{i}}{D_s}\frac{1}{Pe}\nabla\cdot(c_i\nabla \phi) - \frac{D_{i}}{D_s}\frac{1}{Pe}\nabla^2c_i =0,
\end{equation}
where $c_i=C_i/C^\infty_{l}$ are the non-dimensional concentrations and the P\'{e}clet number is defined as $\mathrm{Pe}= U^{ITP}H/D_s$. The charge conservation equation (\ref{eq:ChargeCons}) becomes
\begin{align}
\nabla\cdot&\left[ \left( c_{l} +\frac{D_{s}+D_{0}}{D_{l}+D_{0}}c_{s}+\frac{D_{t}+D_{0}}{D_{l}+D_{0}}c_{t}\right)\nabla\phi\right]\nonumber\\
&=-\left(\frac{z_{l}D_{l}+z_{0}D_{0}}{D_{l}+D_{0}}\right)\nabla^2\left[ c_{l} +\frac{z_{s}D_{s}+z_{0}D_{0}}{z_{l}D_{l}+z_{0}D_{0}}c_{s} +\frac{z_{t}D_{t}+z_{0}D_{0}}{z_{l}D_{l}+z_{0}D_{0}}c_{t}\right].
\end{align}

Under no counterflow, all the zones move with a constant isotachophoretic velocity $U^{ITP}=\mu_{t}E_{t}=\mu_sE_s=\mu_{l}E_{l}$, where $E_i$ ($i=t,s$ or $l$) are the local electric field strengths in the TE, sample or LE zone, respectively, given as
\begin{equation} \label{eq:Efield}
E_i=E_0\frac{ 1/\mu_i}{\sum_i (l_i/\mu_i)},
\end{equation}
where $l_i=L_i/L$. $L_i$ ($i= t,s$ or $l$) is the portion of the channel filled by the respective electrolyte and $L$ is the total length of the channel. $E_0=\Phi/L$ is the average electric field due to the voltage drop $\Phi$ along the channel. Here we have assumed that the sample forms a distinct zone where its concentration remains constant and the overlap region between the zones are negligible. Since in ITP applications the channel is long compared to the section occupied by sample we take the limit $l_s \to 0$ in equation (\ref{eq:Efield}), which then is also valid for the electric field in the TE and LE even in the case of a dispersed sample zone. Also, we assume that the sample zone is situated in the middle of the channel, i.e. we choose $l_{l}=l_{t}=0.5$.

From the conservation of electric current along with electro-neutrality and the negligible diffusive flux at the far-field, a relationship between the TE and LE concentrations can be obtained as
\begin{equation}\label{eq:Kohlrausch}
\frac{C_{t}^\infty}{C_{l}^\infty}=\frac{\mu_{l}+\mu_0}{\mu_{t}+\mu_0}\frac{\mu_{t}}{\mu_{l}},~~
\frac{C_s^\infty}{C_{l}^\infty}=\frac{\mu_{l}+\mu_0}{\mu_s+\mu_0}\frac{\mu_s}{\mu_{l}}
\end{equation}
where the suffix $\infty$ stands for the concentrations far away from the interfaces between the electrolytes \cite{1}$^,$\cite{14}. We will refer to these relations as the Kohlrausch conditions. Note that the second of these relations is only valid in the case where the sample concentration forms a distinct zone with constant concentration, i.e. when it is only marginally affected by diffusive and convective dispersion.

In solving the Nernst-Planck equation, the net ion fluxes through the channel walls are set to zero, i.e. $\textbf{N}_i\cdot\textbf{n}=0$ for $i=t,l$ or $s$, where $\textbf{n}$ is unit outward normal. The electric potential is subjected to insulating boundary condition ($\nabla \phi\cdot\textbf{n}=0$) along the wall. Both the left and right boundaries of the computational domain are placed sufficiently far away from the sample zone such that the distribution of sample ions, $c_s$, is not affected by their presence. The concentrations far away from the transition zones are governed by the Kohlrausch condition (\ref{eq:Kohlrausch}).

Due to the presence of  different electrolytes in different portions of the channel, a sharp change in conductivity will appear across the transition zones. If there is no bulk flow, the width of such a transition zone is of the order of \cite{15}
\begin{equation}
\label{eq:ZoneWidth}
\Delta X= \frac{4 \phi_0}{\Delta E},
\end{equation}
where $\Delta E$ is the change in electric field strength across the zone. Upon application of a voltage drop $\Phi$ across the channel, a locally uniform electric field appears sufficiently far away from each transition zone. This allows for calculation of the electric potential across the channel via charge conservation (\ref{eq:ChargeCons}). In other words, owing to the electro-neutrality assumption we are spared the effort of solving the Poisson equation from which the electric potential is usually derived.

The electroneutrality assumption\cite{13} relies on the fact that its inevitable violation dictated by a changing electric field due to Gauss's law, $\rho_e=\varepsilon_e \mathrm{div} \mathbf{E}$, is much smaller than the total amount of charges present, $\sim F|z_i|c_i$. For ITP applications the ratio between the two is typically of the order of $10^{-4}-10^{-5}$  (cf. supplementary material for \cite{Baier_2011, 22}) which allows us to safely neglect the impact of $\rho_e/F$ on the concentrations. However, this is not the only place where the charge density enters, since in terms of Maxwell stresses it also contributes as a body force, $\sim \rho_e \mathbf{E}$, in the momentum equation for the fluid \cite{Lin_2004, 22}. This issue of Maxwell stresses in the context of ITP  in a similar setting was discussed in \cite{Baier_2011}, where it is concluded that the body force term becomes important for large applied fields with very narrow transition zones and hence large field gradients, something that is not achieved with the strong convective dispersion considered here.

In order to make quantitative predictions using this model, the electrophoretic mobilities of all species, concentration of the leading electrolyte, voltage drop across the channel, and depth of the channel must be specified. Note that the TE concentration can be evaluated by using the Kohlrausch conditions (\ref{eq:Kohlrausch}). The electric field strengths at the ends of the channel can be adjusted via relation (\ref{eq:Efield}). We use $\varepsilon_e = 78.5\times 8.85\cdot 10^{-12}\,\mathrm{C/(Vm)}$, $T=300\,\mathrm{K}$ and $\phi_0=25.9\,\mathrm{mV}$.

\section{Numerical Methods}

We developed a computer code to compute the governing equations based on the numerical algorithm as described below. The non-dimensional equations for ion transport and electric field are computed in a coupled manner through the  finite volume method \cite{17}. The computational domain is subdivided into a number of control volumes. When the electromigration in the Nernst-Planck equations dominates the electro-diffusional transport of ions, the transport equations show hyperbolic characteristics. Due to the hyperbolic nature of the ion transport equations, we adopt a higher-order upwind scheme QUICK\cite{18} (Quadratic Upwind Interpolation Convection Kinematics) to discretize the electromigration and convection terms in the ion transport equations. The diffusion flux at the control volume interfaces is estimated by a linear interpolation of variables between the two neighbors to either sides of the control volume interfaces. Details of the spatial discretisation scheme can be found in the supplementary material. An implicit first-order scheme is used to discretize the time derivatives present. At every time level we solve the ion transport equations iteratively, as the ion transport equations and the charge conservation equation are coupled. The iteration procedure starts with a guess for the electric potential at each cell center. At every iteration, the elliptic PDE for the charge conservation equation is integrated over each control volume through the finite volume method. The discretized equations are solved by a line-by-line iterative method along with the successive-over-relaxation (SOR) technique. The iterations are continued until the absolute difference between two successive iterations becomes smaller than the tolerance limit $10^{-6}$ for concentration as well as for the electric potential. The details on discretization of the governing equations are provided in the supplementary material.

A steady state solution is achieved by taking a sufficient number of time steps until the concentration distributions remain unchanged with time. The initial condition for ITP with counterflow is governed  by the solution of the corresponding ideal ITP case (without convection). The solution for ideal ITP is also obtained based on the algorithm as described above and the parabolic velocity profile is imposed after a steady isotachophoretic zone is formed. We find that the steady-state state is archived after a short transition phase. The time evolution and formation of steady-state is illustrated in figure \ref{fig:time} of the supplementary material. However, in this paper all results presented correspond to the situation where a steady state has been reached.

As the variables within the transition zone between two electrolytes change more rapidly than elsewhere, a nonuniform grid spacing along the $x$-direction and uniform spacing along $y$-direction is chosen. The grid size $\delta x$ within the transition zones is relatively small and is increased with a constant increment as we move away from the transition zone. The smallest $\delta x$ is chosen such that the Courant-Friedrich-Levy (CFL) criterion is satisfied. For the present computations, $\delta x$ is taken to be $0.005$ for the finest grid and it is $0.01$ for coarse grids. Here $\delta y$ is taken to be $0.005$ and $\delta t=0.001$. A grid independency test and validation of our algorithm by comparing with the analytical solution for ideal ITP (Goet et al. \cite{15}) is presented in figure \ref{fig:grid} in the supplementary material. The efficiency of the present numerical code in resolving the transient zones for peak and plateau-mode ideal ITP in steady-state is illustrated in figure \ref{fig:pureITP_varSample} of the supplementary material.

\section{Taylor-Aris Dispersion model}

To analyze the dispersion effect on the sample when a counterflow is applied, we consider a one-dimensional model based on Taylor-Aris dispersion \cite{19}$^,$\cite{20}. Taylor's original work was concerned with dispersion solely due to the pressure driven flow through a channel. In that case the mechanism at work is the interplay of convective dispersion and lateral diffusion. Convective dispersion alone would result in a dispersion linear with time, but its effect is limited by the lateral diffusion over the cross section of the channel. Combined, these phenomena result in a dispersion proportional to the square root of time, such that the area-averaged sample distribution can be considered to spread under the influence of an effective axial diffusion. For two reasons it is a priori unclear whether such a Taylor-Aris dispersion model can be applied to the situation considered in this article. First, electromigration appears as an additional transport process that results in a sharpening of the sample distribution. Secondly, an assumption implicit to the Taylor-Aris model is the infinite time limit, i.e. the model is valid only on time scales large compared to the time a molecule takes to diffusively sample the channel cross section. In turn, this means that it is usually applicable only to comparatively broad (in axial direction) sample distributions with $\sigma/H \gg \mathrm{Pe}$, where $\sigma$ is the length scale of a transition zone. For this reason one would a priori not expect that a description based on an effective axial diffusivity is applicable. Nevertheless, such models have been successfully used previously in similar situations \cite{Saville_1990}$^,$\cite{21}$^,$\cite{22}. Therefore we will investigate the applicability of this model to the case considered here.

The core of the Taylor-Aris dispersion model is the effective axial diffusion due to convection. In particular, the diffusion coefficient is replaced by a term dependent on the P\'eclet number squared
\begin{equation}
\label{eq:EffDiff}
D_{\mathrm{eff},i}=D_i\left[1+ \beta\left(\frac{H  U^{ITP}}{D_i}\right)^2\right],
\end{equation}
where $i=t,s$ or $l$. For a parallel-plates channel the value of the constant $\beta$ is $1/210$. The dispersion model is based on the area-averaged ion concentrations
\begin{equation}
\overline{C}_i(X)=\frac{1}{H} \int_0^H dY\,C_i(X,Y).
\end{equation}
Likewise, we write area-averaged transport equations for the ion concentrations in which the diffusivity is replaced by the effective axial diffusivity according to Taylor and Aris
\begin{equation}
\label{eq:TaylorAris}
\frac{\partial \overline{C}_i }{\partial t}+ \frac{\partial }{\partial X} \left[ \left( z_i \mu_i E_x~ \overline{C}_i  - D_{\mathrm{eff},i}\frac{\partial \overline{C}_i }{\partial X}\right) -U^{ITP} ~\overline{C}_i \right] = 0.
\end{equation}
The axial electric field is determined by solving the corresponding one-dimensional charge conservation equation assuming electro-neutrality throughout
\begin{equation}\label{eq:ChargeConsAris}
\frac{\partial}{\partial X}(\overline{\nu} E_x)=F\sum_i{D_i z_i \frac{\partial^2\overline{C}_i}{\partial X^2}},
\end{equation}
where $\overline{\nu}$ is the area-averaged conductivity. The equation for the electric field is coupled with the transport equations (\ref{eq:TaylorAris}). The numerical solution of the species transport equations above is achieved by employing a higher-order accurate upwind differencing scheme. The equation for the electric field is solved along with the ion transport equations in a coupled manner as outlined before.

\section{Results and discussion}\label{sec:Results}

We consider the LE and common ion diffusivities as fixed values throughout this study, setting $D_{l}=D_0=7.0\times 10^{-10}\,\mathrm{m^2/s}$. The diffusivities of the TE and sample ions are prescribed by specifying the two non-dimensional parameters $k_1=D_{l}/D_{t}$ and $k_2=D_{l}/D_s$. The corresponding mobilities are related to the diffusivities via the Nernst-Einstein relation, $\mu_i=D_i F/RT$. Since the sample mobility lies between the mobilities of TE and LE, these parameters can be varied subject to the restriction $k_1>k_2>1$. The scale of the applied electric field is set to $E_0 = 10^5\,\mathrm{V/m}$ from which the field in the respective buffers can be inferred using equation (\ref{eq:Efield}). The bulk LE concentration is taken as $C^\infty_{l}=10^{-3}\,\mathrm{M}$, so that the Debye layer thickness ($\sim 10\,\mathrm{nm}$) is much smaller than the depth of the channel considered here ($>5\,\mathrm{\mu m}$). The bulk TE concentration can be inferred from the Kohlrausch condition (\ref{eq:Kohlrausch}), which in our example becomes $C^\infty_{t}/C^\infty_{l}=2/(1+k_1)$. A corresponding upper bound can be obtained for the maximum of the sample concentration, which due to dispersion will usually be much smaller in the cases considered here, however.

In ITP experiments one considers two modes of operation, peak and plateau mode, depending on the amount of sample present in the system. "Plateau mode" refers to the case where the sample-zone is wide compared to the transition zones, i.e. the sample concentration distribution forms a constant plateau with blurred boundaries towards LE and TE. "Peak mode" refers to the other extreme of a very short sample zone, where the two transition zones at both sides of the sample overlap or more precisely, when the sample is entirely within the diffused interface between LE and TE. In this situation the influence of the analytes on the overall conductivity is typically negligible. Everything else being equal, the difference between peak and plateau mode lies in the amount of sample present. In the present case, owing to the strong additional dispersion due to convection, a larger amount of sample will still lead to a pronounced peak instead of a plateau, as will be elaborated below. In order to distinguish this regime from peak mode as observed without convective dispersion, we will call this situation ``dispersed plateau mode". This is the regime that we will mainly, but not exclusively, consider presently. It is different from plateau mode in that the area averaged sample distribution shows a pronounced peak and differs from peak mode in that the sample strongly affects the overall conductivity.

In order to obtain comparable results for the distortion of the sample region due to the applied Poiseuille flow, we fix the area-averaged amount of sample
\begin{equation}\label{eq:SampleAmount}
{\cal C}_s %= \frac{1}{H} \int_0^H dY \int_{-\infty}^\infty dX \, C_s(X,Y) 
		= \int_{-\infty}^{\infty} dX \, \bar C_s(X)
\end{equation}
present in the channel. In that way the area-averaged distribution without convective dispersion is independent of the channel height. Note that the exact form of the distribution will depend on the choice of other parameters such as $k_1$ and $k_2$ due to the different concentrations dictated by the Kohlrausch conditions (\ref{eq:Kohlrausch}) or the width of transition zones (\ref{eq:ZoneWidth}). Unless stated otherwise we will mainly use the values of ${\cal C}_{s,0}=40\,\mathrm{\mu mol/m^2} = 40\,\mathrm{\mu m}\cdot C^\infty_{l}$ and ${\cal C}_{s,0}=4\,\mathrm{\mu m}\cdot C^\infty_{l}$ throughout; the former value leads to a distinct plateau without convective dispersion while it represents dispersed plateau mode for wide enough channels, while the latter corresponds to a sample distribution in the form of a Gaussian peak without convective dispersion, cf. figure \ref{fig:pureITP_varSample} in the supplementary material.

We remark that without convective dispersion the width of the sample zone in plateau mode can easily be inferred from the Kohlrausch condition (\ref{eq:Kohlrausch}) and the amount of sample present, neglecting the width (\ref{eq:ZoneWidth}) of the diffusive zones, to $L_s\approx{\cal C}_s/C_s^\infty={\cal C}_s/C_l^\infty\cdot (1+k_2)/2$, which gives a baseline for the width of the sample distribution.

%%%%%%%%%%%%%%%%%%%%%%%%%%%%%%%%%%
\begin{figure}
\subfigure[]{\label{c11}\includegraphics[height=1.08345 in]{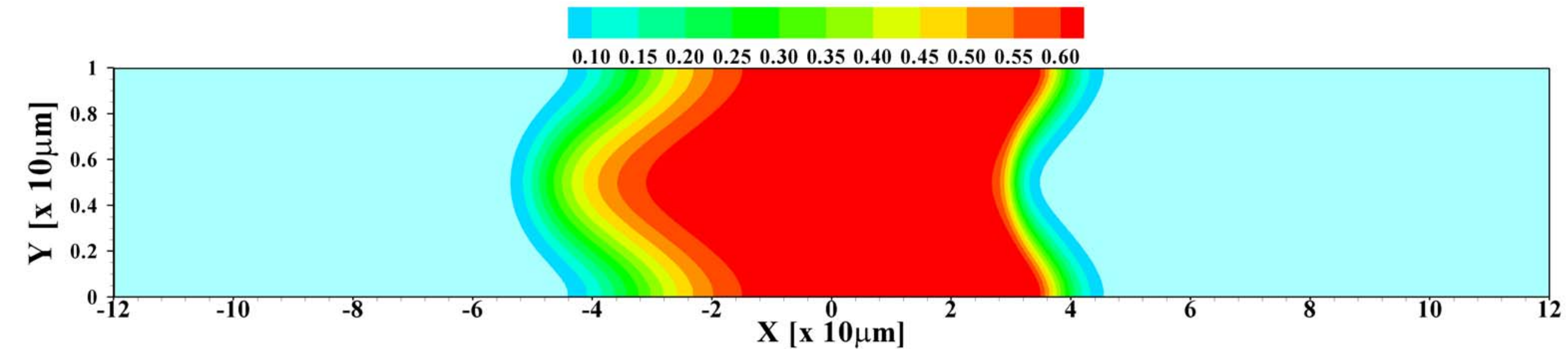}}
\subfigure[]{\label{c12}\includegraphics[height=1.7 in]{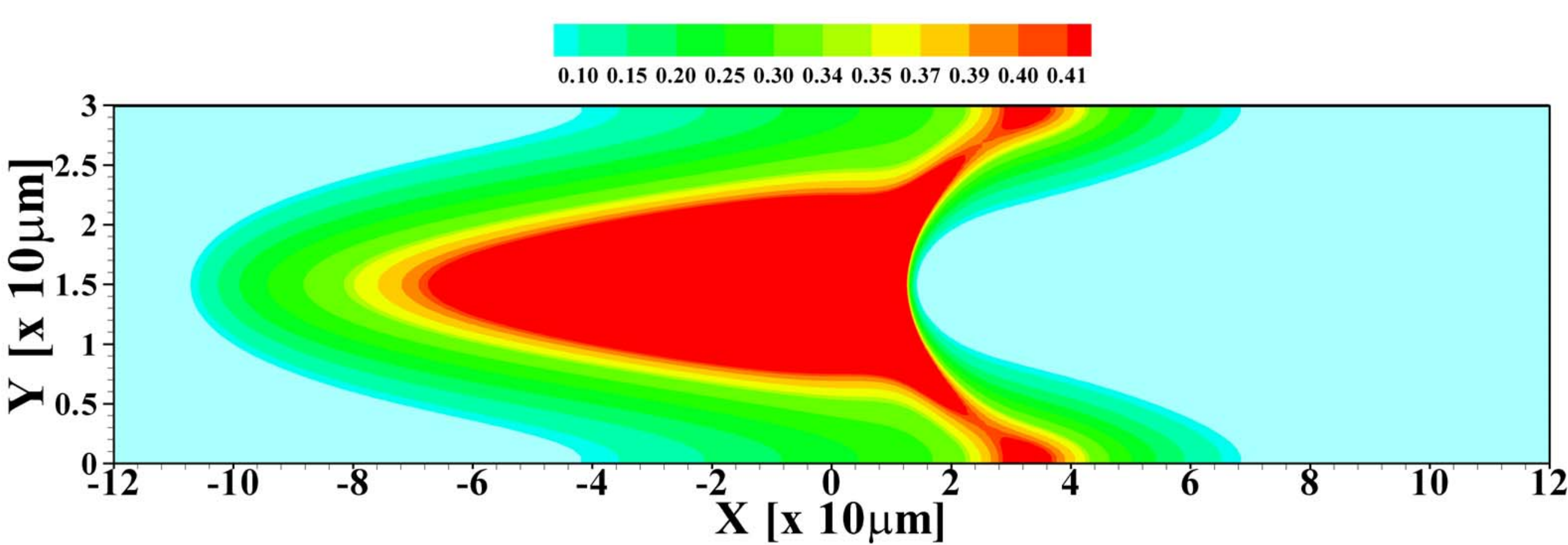}}
\caption{\label{fig:VariableH_2D} Sample concentration obtained in the 2D model for $k_1(=\mu_{l}/\mu_{t}=D_{l}/D_{t})=3$, $k_2(=\mu_{l}/\mu_s=D_{l}/D_s)=2$ with channel depths of (a) $H=10 \,\mathrm{\mu m}$ ($\mathrm{Pe}=40$) and (b) $H=30\,\mathrm{\mu m}$ ($Pe=120$). The amount of sample is ${\cal C}_s/C^\infty_{l} = 40\,\mathrm{\mu m}$. Both coordinates axes show the length in units of $10^{-5} m$, but were rescaled differently for better visualization of the sample distribution.}
\end{figure}

\begin{figure}
\includegraphics[height=2.5in]{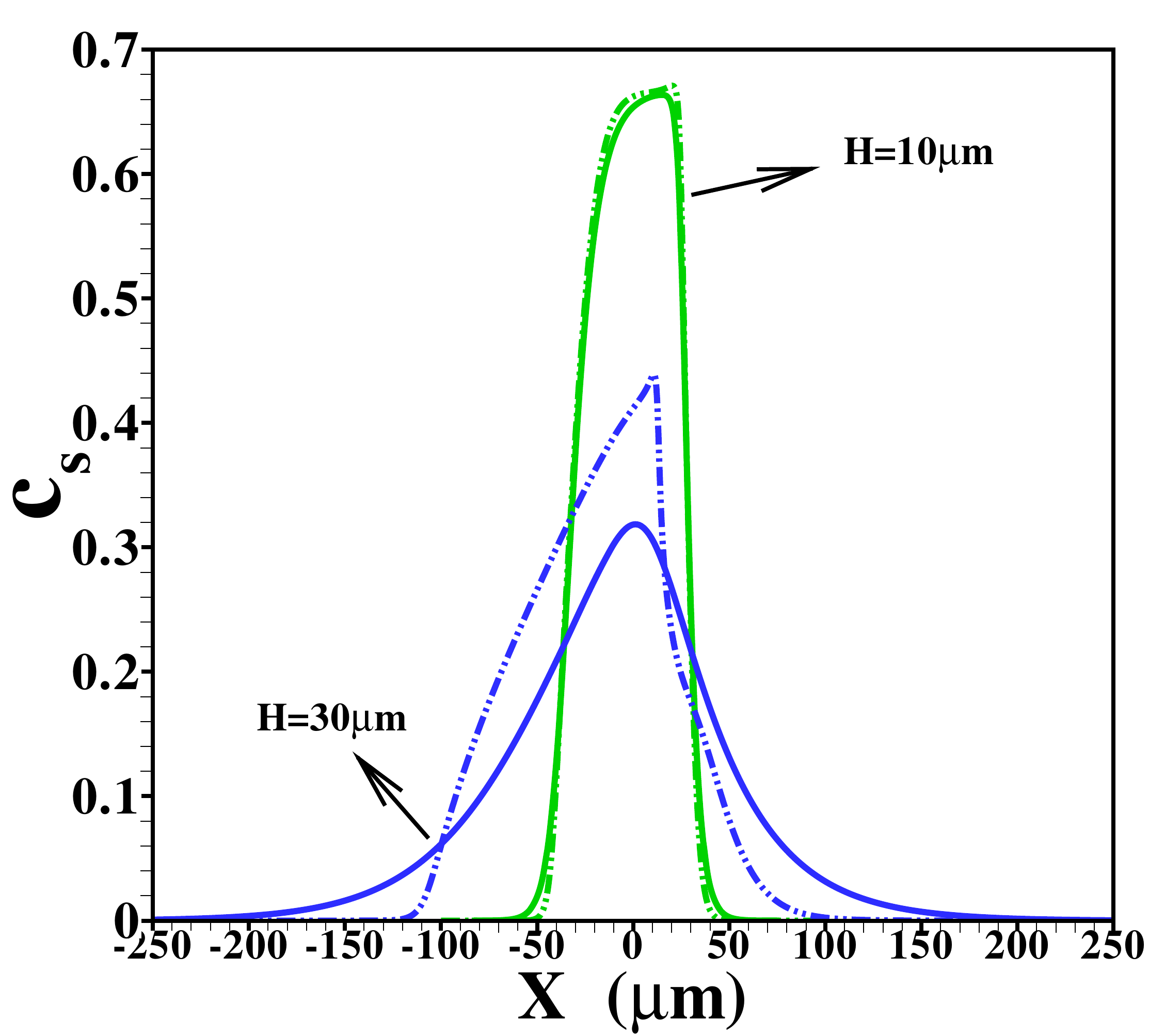}
%\subfigure[][$\;H=30\,\mathrm{\mu m}$]{\label{c22}\includegraphics[height=2.5in]{2b}}\hspace{.25cm}
\caption{\label{fig:VariableH_1D} Effect of the channel height on the area-averaged sample concentration profile and comparison with the corresponding profile obtained in the Taylor-Aris dispersion model for fixed values of the diffusivity ratios i.e., $k_1(=\mu_{l}/\mu_{t}=D_{l}/D_{t})=3$ and  $k_2(=\mu_{l}/\mu_s=D_{l}/D_s)=2$ when the depth of the channel is chosen as $H=10\,\mathrm{\mu m}$ and {$H=30\,\mathrm{\mu m}$}, respectively. The amount of sample is ${\cal{C}}_s/C_{l}^\infty=40\,\mathrm{\mu m}$.}
\end{figure}
%%%%%%%%%%%%%%%%%%%%%%%%%%%%%%%%%%

In figure \ref{fig:VariableH_2D} the corresponding sample concentration profiles are shown for two different channel heights. As mentioned, all results correspond to a situation where a steady state has been reached. Figure \ref{fig:VariableH_1D} shows the corresponding area-averaged sample concentrations, both derived from the 2D model and from the 1D Taylor-Aris dispersion model (\ref{eq:TaylorAris}) with effective diffusivity (see supporting information for a combined graph of the TE, sample and LE distributions in this situation). From these figures it becomes apparent that whether a plateau or peak is obtained now becomes dependent on the channel height. For the shallow channel a plateau is observable in the area-averaged concentrations, contrasting the distinct peak observed for the wider channel, characteristic for the ``dispersed plateau mode''. Since in both cases the ITP velocity is the same, the larger dispersion is a direct consequence of the larger P\'{e}clet number for the wider channel. We also note that the 1D model based on the effective Taylor-Aris diffusion captures the area-averaged distribution quite well. This is particularly true for the narrower channel. In the wider channel the larger discrepancy is due to the slight focusing of the sample in the center region of the channel, which effectively results in a narrower peak than determined from the 1D model. We will further investigate this below by varying the total sample concentration present in the channel.
\begin{figure}
\subfigure[]{\label{c31}\includegraphics[height=3in]{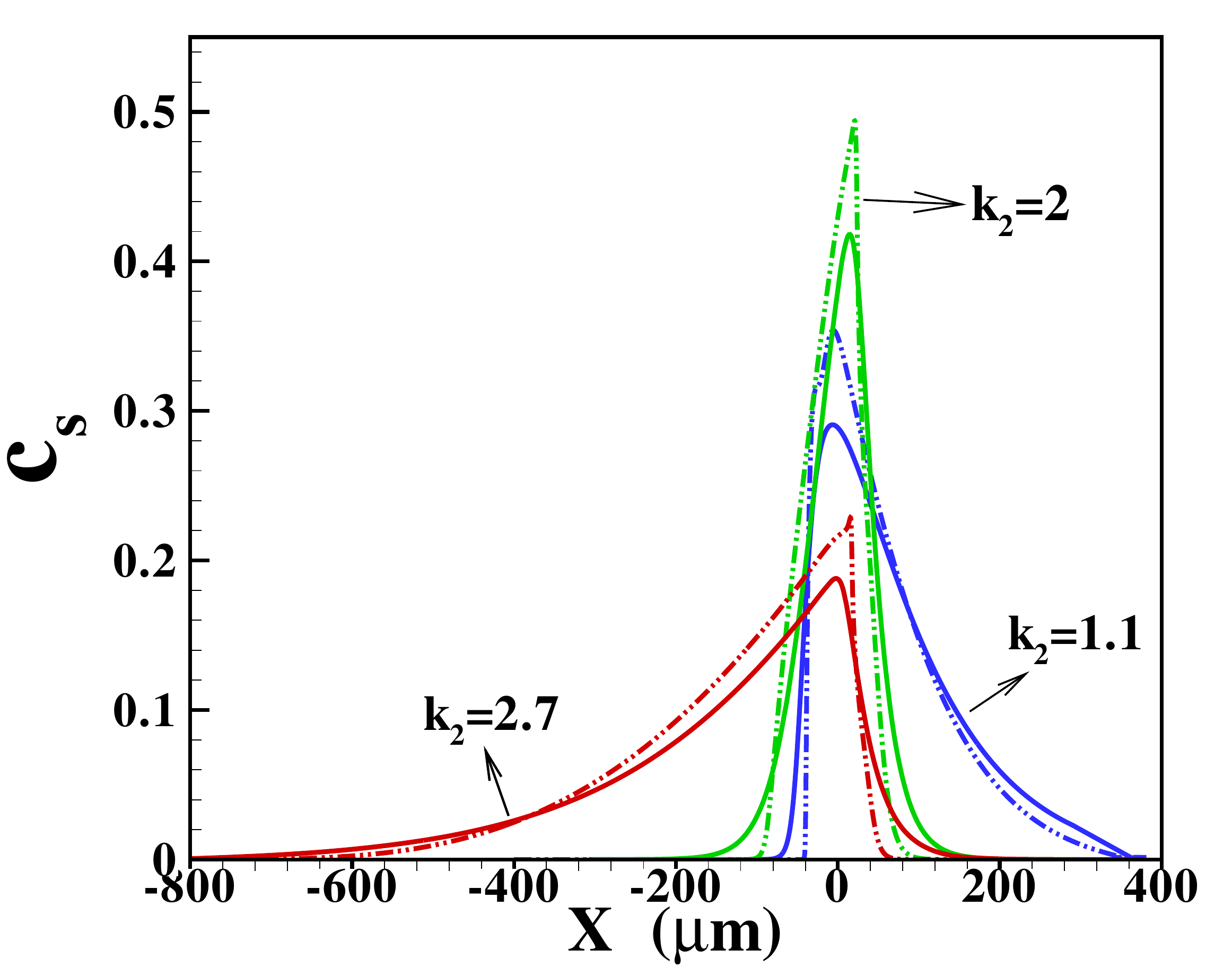}}\hspace{.25cm}
\subfigure[]{\label{c32}\includegraphics[height=3in]{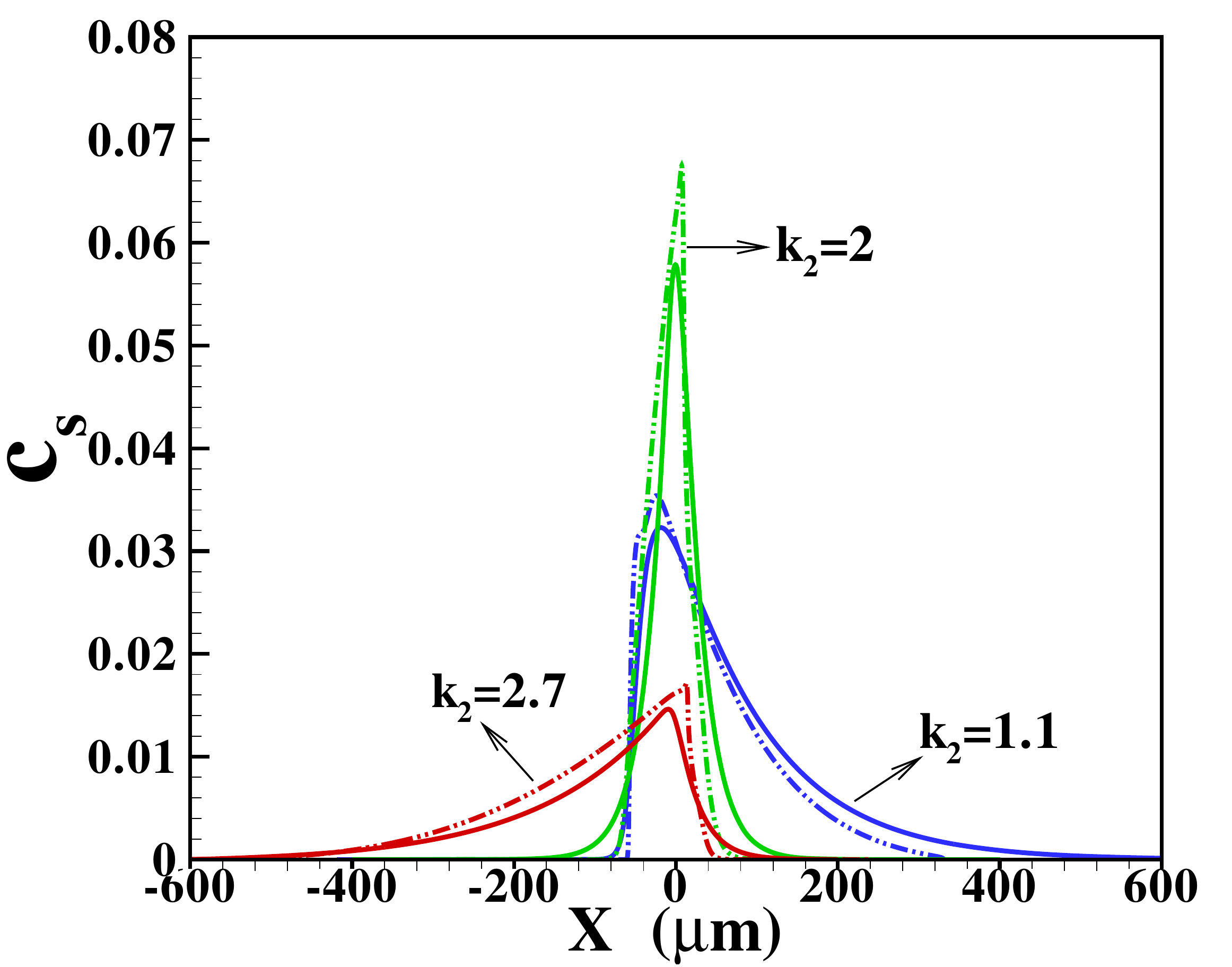}}\hspace{.25cm}
\caption{\label{fig:Variable_k2} Area-averaged sample concentration profile obtained in the 2D simulation for different values of the diffusivity of sample electrolyte i.e., $k_2(=\mu_{l}/\mu_s=D_{l}/D_s)=1.1$ ($\mathrm{Pe}=55$), $k_2=2$ ($\mathrm{Pe}=100$), $k_2=2.7$ ($\mathrm{Pe}=135$). Here the LE to TE diffusivity ratio i.e, $k_1(=\mu_{l}/\mu_{t}=D_{l}/D_{t})=3$ and channel depth is $H=25\,\mathrm{\mu m}$. The amount of sample is (a) ${\cal C}_s/C^\infty_{l} = 40\,\mathrm{\mu m}$; (b) ${\cal C}_s/C^\infty_{l} = 4\,\mathrm{\mu m}$. A comparison with the corresponding profile obtained in the Taylor-Aris dispersion model is also made. Dashed lines represents the results obtained from the 2D model and the solid line represents corresponding results from the 1D model.}
\end{figure}

Before doing so, we investigate the area-averaged distributions obtained with the two models for different ratios of the diffusivities, cf. figure~\ref{fig:Variable_k2}~(a) when the sample amount is ${\cal C}_s/C^\infty_{l} = 40\,\mathrm{\mu m}$.  Corresponding results for the reduced sample amount i.e., ${\cal C}_s/C^\infty_{l} = 4\,\mathrm{\mu m}$ is shown in figure~\ref{fig:Variable_k2}~(b). In the cases where the sample diffusivity is very close to the diffusivity for either LE or TE one expects a very wide transition zone between the sample and the respective electrolyte, since hardly any electromigrative sharpening is present. On the other hand, for the transition towards the electrolyte with markedly different diffusivity a much narrower zone is expected. This is indeed seen in figure~\ref{fig:Variable_k2}~(a) and (b). We remark that the three situations depicted correspond to different P\'{e}clet numbers, since the parameter varied is the sample diffusivity. Nevertheless, the total width of the sample zone is affected more strongly by the long tails towards the electrolyte with similar diffusivity than by the P\'{e}clet number. We will further investigate this fact below. Also note that again the Taylor-Aris model quite successfully captures the area-averaged concentration profiles, in particular towards the electrolyte with similar diffusivity. The sharper profile towards the electrolyte with dissimilar diffusivity in the 2D compared with the 1D case is again due to the more detailed 2D structure with a slight focusing in the channel center emerging in this model. Comparing figure~\ref{fig:Variable_k2}~(a) and (b) we find that in the dispersed plateau mode the profiles for the sample distribution are largely independent of its amount.

%%%%%%%%%%%%%%%%%%%%%%%%%%%%%%%%%%
\begin{figure}
\subfigure[]{\label{c41}\includegraphics[height=1.5 in]{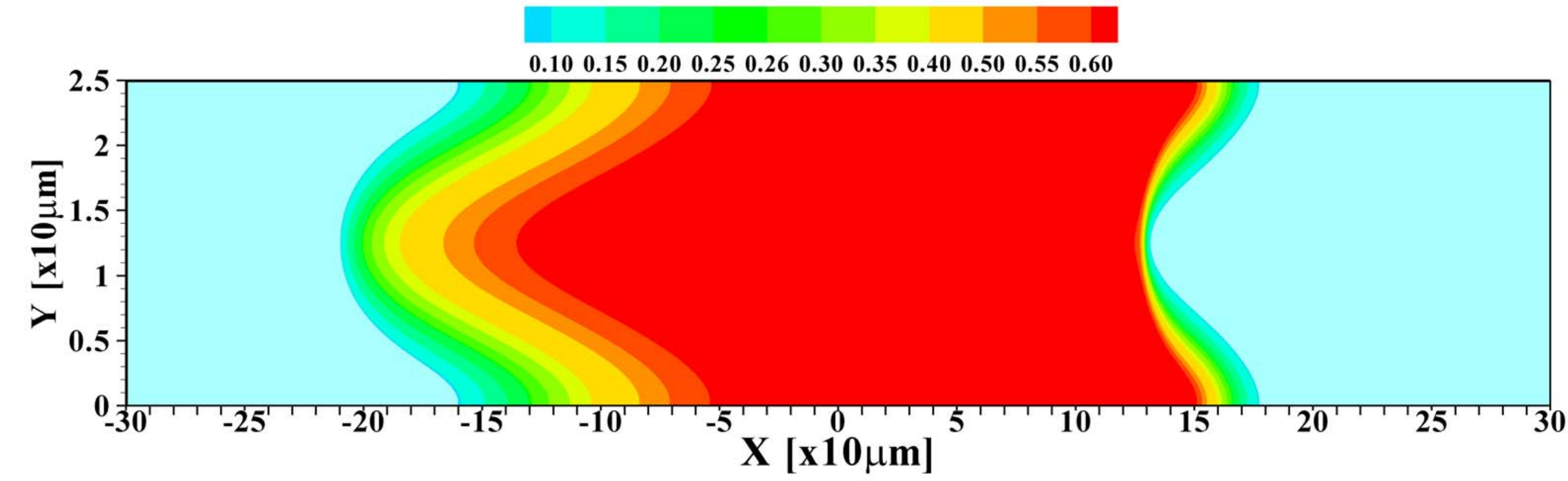}}
\subfigure[]{\label{c42}\includegraphics[height=1.5 in]{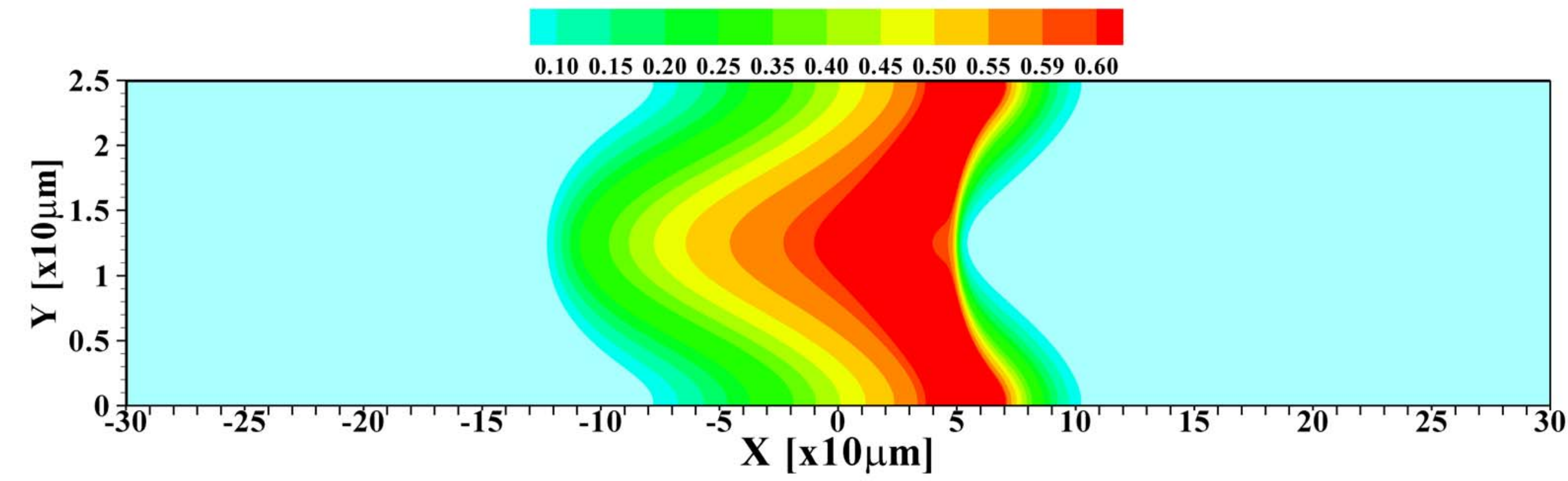}}
\subfigure[]{\label{c43}\includegraphics[height=1.5 in]{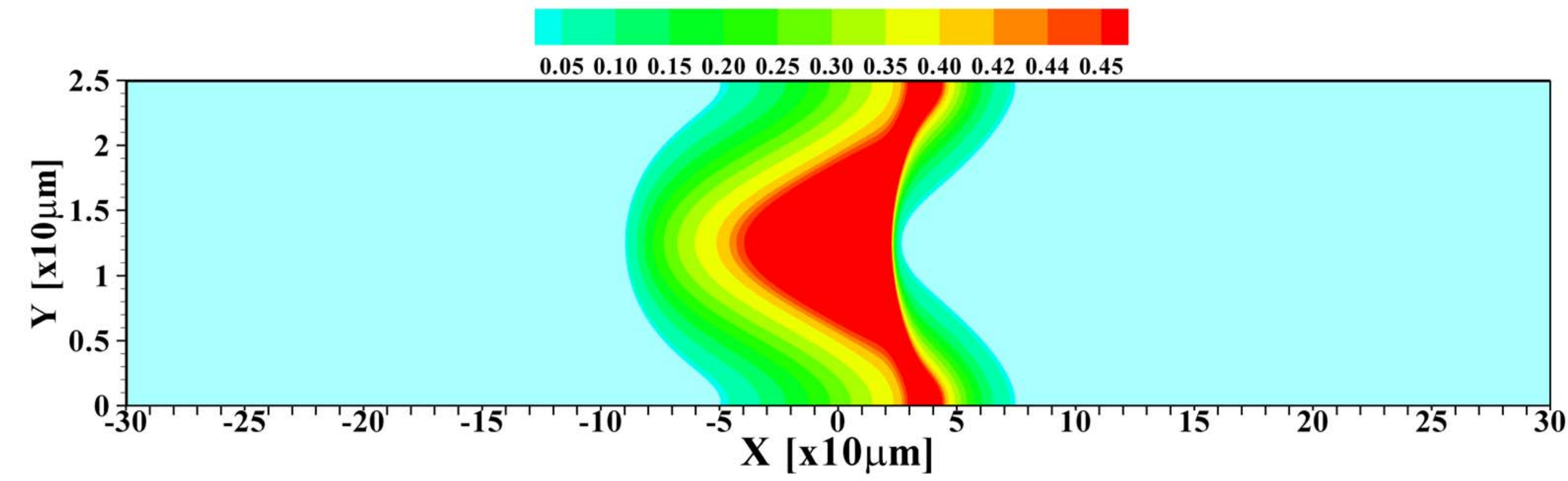}}
\subfigure[]{\label{c44}\includegraphics[height=1.5 in]{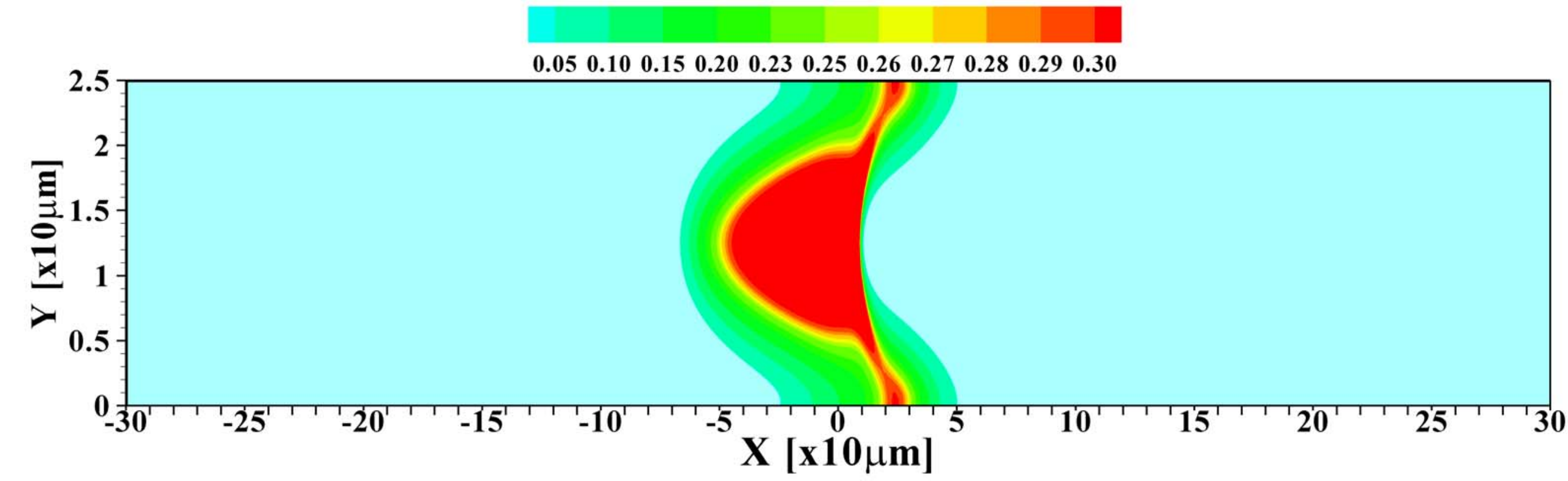}}
\caption{\label{fig:Variable_ms_2D} Sample concentration obtained in the 2D model for $k_1(=\mu_{l}/\mu_{t}=D_{l}/D_{t})=3$ and  $k_2(=\mu_{l}/\mu_s=D_{l}/D_s)=2$ when (a) ${\cal C}_s=5\, {\cal C}_{s,0}$, (b) ${\cal C}_s=2\, {\cal C}_{s,0}$, (c) ${\cal C}_s=1\, {\cal C}_{s,0}$ and (d) ${\cal C}_s=0.5\, {\cal C}_{s,0}$ with ${\cal C}_{s,0}/C^\infty_{l} = 40\,\mathrm{\mu m}$. The channel depth is $H=25\,\mathrm{\mu m}$. Note that both the coordinate axes are multiplied by a factor $10^{-5} m$.}
\end{figure}

\begin{figure}
\subfigure{\label{c51}\includegraphics[height=3.5 in]{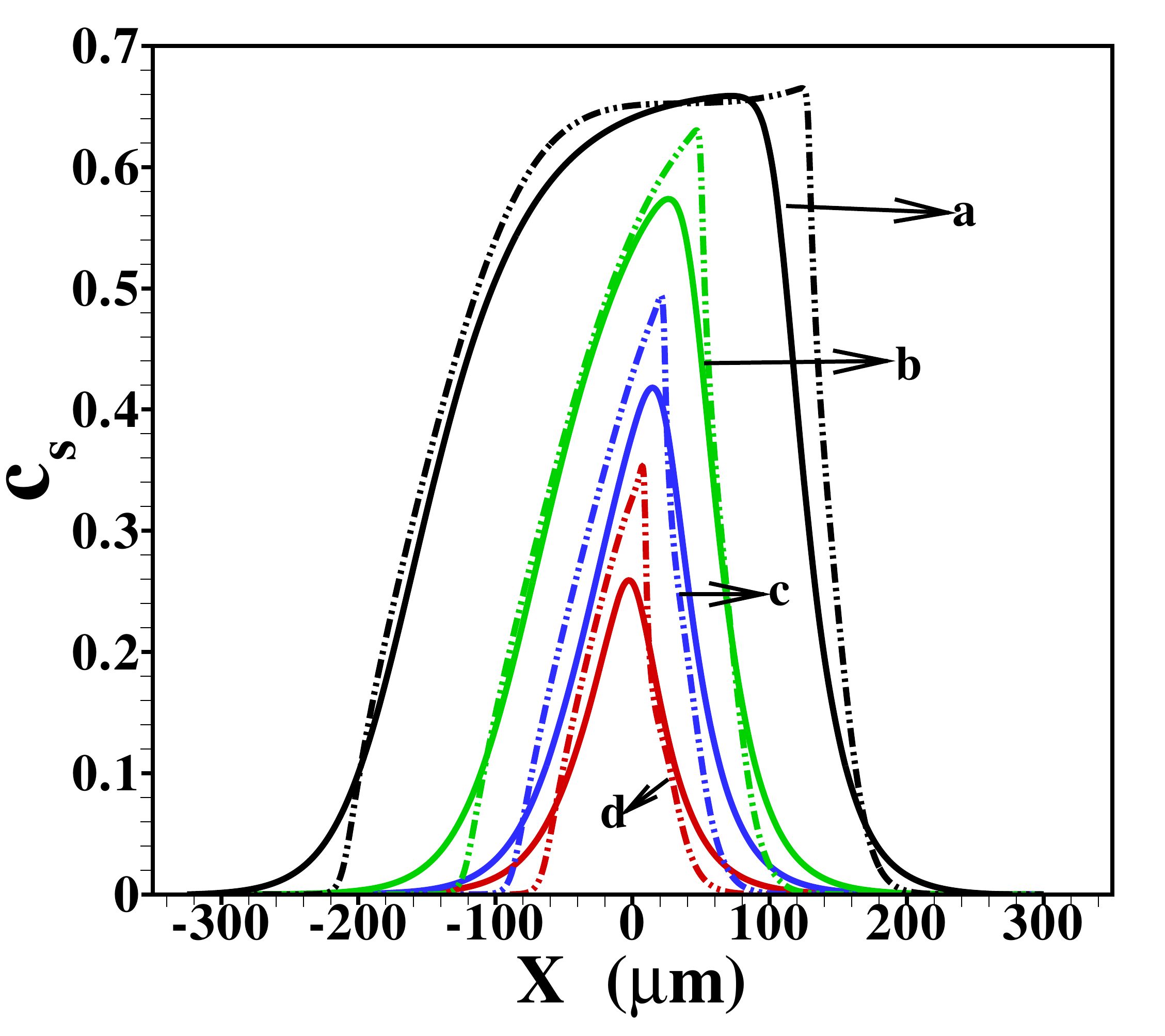}}\hspace{.25cm}
\caption{ \label{fig:Variable_ms} Comparison of the area-averaged sample concentration profiles obtained in the 2D simulation with the corresponding profile obtained in the Taylor-Aris dispersion model for fixed values of $k_1(=\mu_{l}/\mu_{t}=D_{l}/D_{t})=2$ and  $k_2(=\mu_{l}/\mu_s=D_{l}/D_s)=1.5$ when (a) ${\cal C}_s=5\, {\cal C}_{s,0}$, (b) ${\cal C}_s=2 {\cal C}_{s,0}$, (c) ${\cal C}_s=1 {\cal C}_{s,0}$ and (d) ${\cal C}_s=0.5\, {\cal C}_{s,0}$  with ${\cal C}_{s,0}/C^\infty_{l} = 40\,\mathrm{\mu m}$. The depth of the channel is $H=25\,\mathrm{\mu m}$. The solid line represents the results obtained from 1D model and dashed lines from 2D model.}
\end{figure}
%%%%%%%%%%%%%%%%%%%%%%%%%%%%%%%%%%

To further investigate the validity of the 1D model, the total amount of sample present in the channel is varied. Figure \ref{fig:Variable_ms_2D} shows the 2D profiles obtained for amounts ranging from $1/2$ to $5$ times ${\cal C}_{s,0}$, the amount considered so far. For the smallest amount of sample, one again observes the focusing of the sample in the central region of the channel. Contrasting that, for the largest amount of sample present, the situation may be considered as plateau mode. The corresponding area-averaged concentration profiles are shown in figure \ref{fig:Variable_ms}, both for the 2D and 1D cases. Again, the 1D model is surprisingly effective in predicting the profiles, with the quality of the predictions deteriorating for small sample amounts due to the strong 2D structure of the profile, leading to a sharper focusing than predicted by the 1D model.

In order to quantify the dispersion, we define the width of the sample zone as the second moment of the sample distribution
\begin{equation}
\sigma^2 %= \frac{1}{H{\cal C}_s} \int_0^H dY \int_{-\infty}^\infty dX\, (X^2-\bar{X}^2) \,C_s(X,Y)
  =\frac{1}{{\cal C}_s} \int_{-\infty}^\infty dX\, (X^2-\bar{X}^2)\,\bar C_s(X),
\end{equation}
where the center of the distribution is defined as
\begin{equation}
\bar{X} %= \frac{1}{H{\cal C}_s}\int_0^H dY \int_{-\infty}^\infty dX\, X\,C_s(X,Y)
=\frac{1}{{\cal C}_s} \int_{-\infty}^\infty dX\, X\,\bar C_s(X).
\end{equation}
Additionally, we also measure the skewness of the distribution through the third normalised moment as
\begin{equation}
\gamma_1 %= \frac{1}{H{\sigma^3\cal C}_s} \int_0^H dY \int_{-\infty}^\infty dX\, (X -\bar{X} )^3 \,C_s(X,Y)
  =\frac{1}{{\cal C}_s} \int_{-\infty}^\infty dX\, \left(\frac{X -\bar{X}}{\sigma} \right)^3 \,\bar C_s(X),
\end{equation}
to further quantify the similarity between results for the sample distribution obtained within the 2D model and the 1D Taylor-Aris dispersion model. The value of skewness may be positive, negative or zero depending on whether the sample electrolyte is skewed to the right (positive skew), to the left (negative skew) or symmetric (zero skew).

%%%%%%%%%%%%%%%%%%%%%%%%%%%%%%%%%%
\begin{figure}
\subfigure[]{\label{c71}\includegraphics[height=3in]{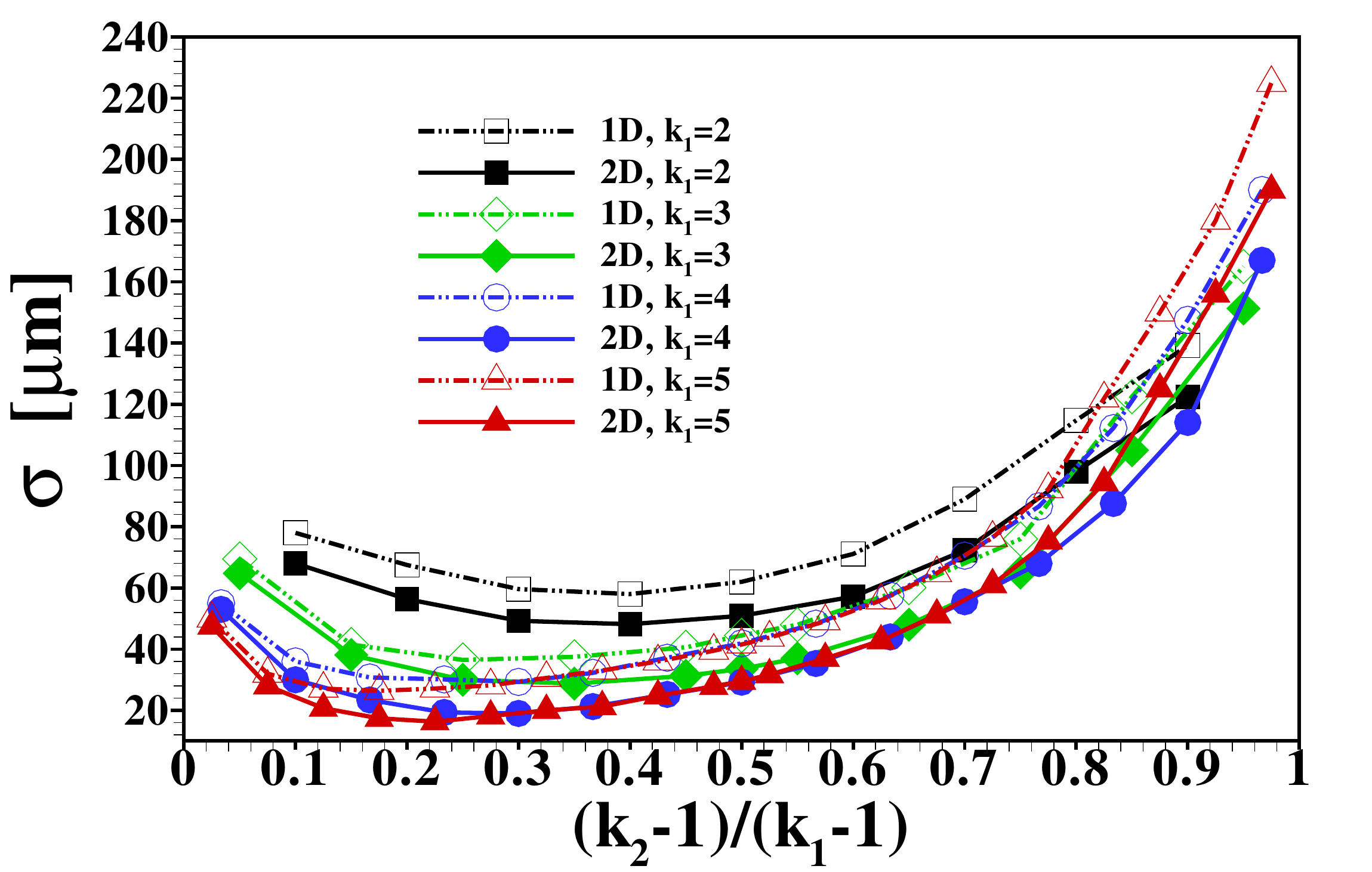}}\hspace{.25cm}
\subfigure[]{\label{c72}\includegraphics[height=3in]{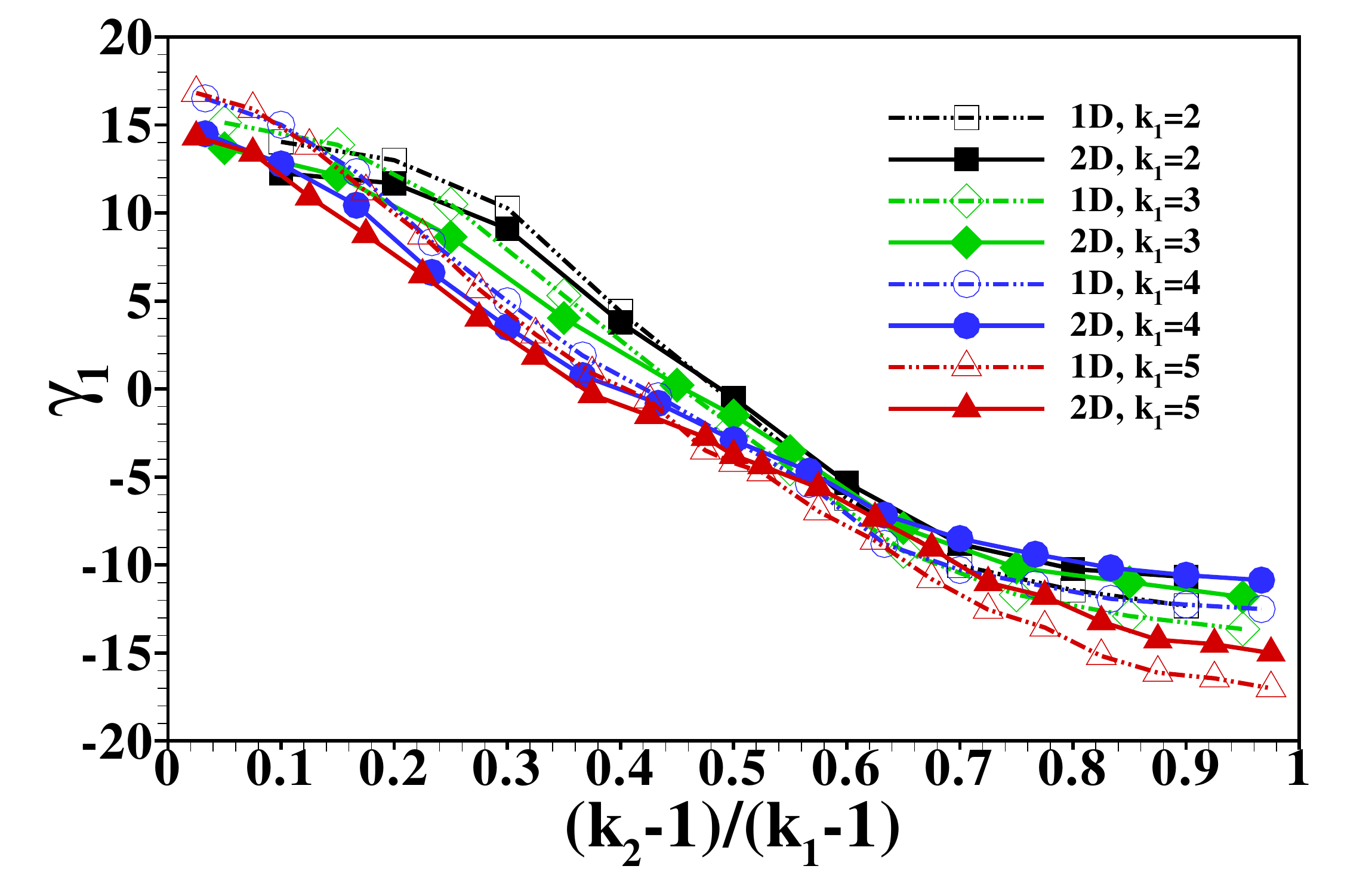}}\hspace{.25cm}
\caption{\label{fig:StDev_allk}Variation of the (a) standard deviation and (b)  skewness of the sample distribution with $k_2(=\mu_{l}/\mu_s=D_{l}/D_s)$ for a fixed value of $k_1(=\mu_{l}/\mu_{t}=D_{l}/D_{t})=2$, $3$, $4$ and $5$.  The amount of sample is ${\cal{C}}_s/C_{l}^\infty=40\,\mathrm{\mu m}$. Lines with unfilled symbols represent the results from the 1D model (Taylor-Aris model), lines with filled symbols those from the 2D model.}
\end{figure}

\begin{figure}
\subfigure[]{\label{c61}\includegraphics[height=3in]{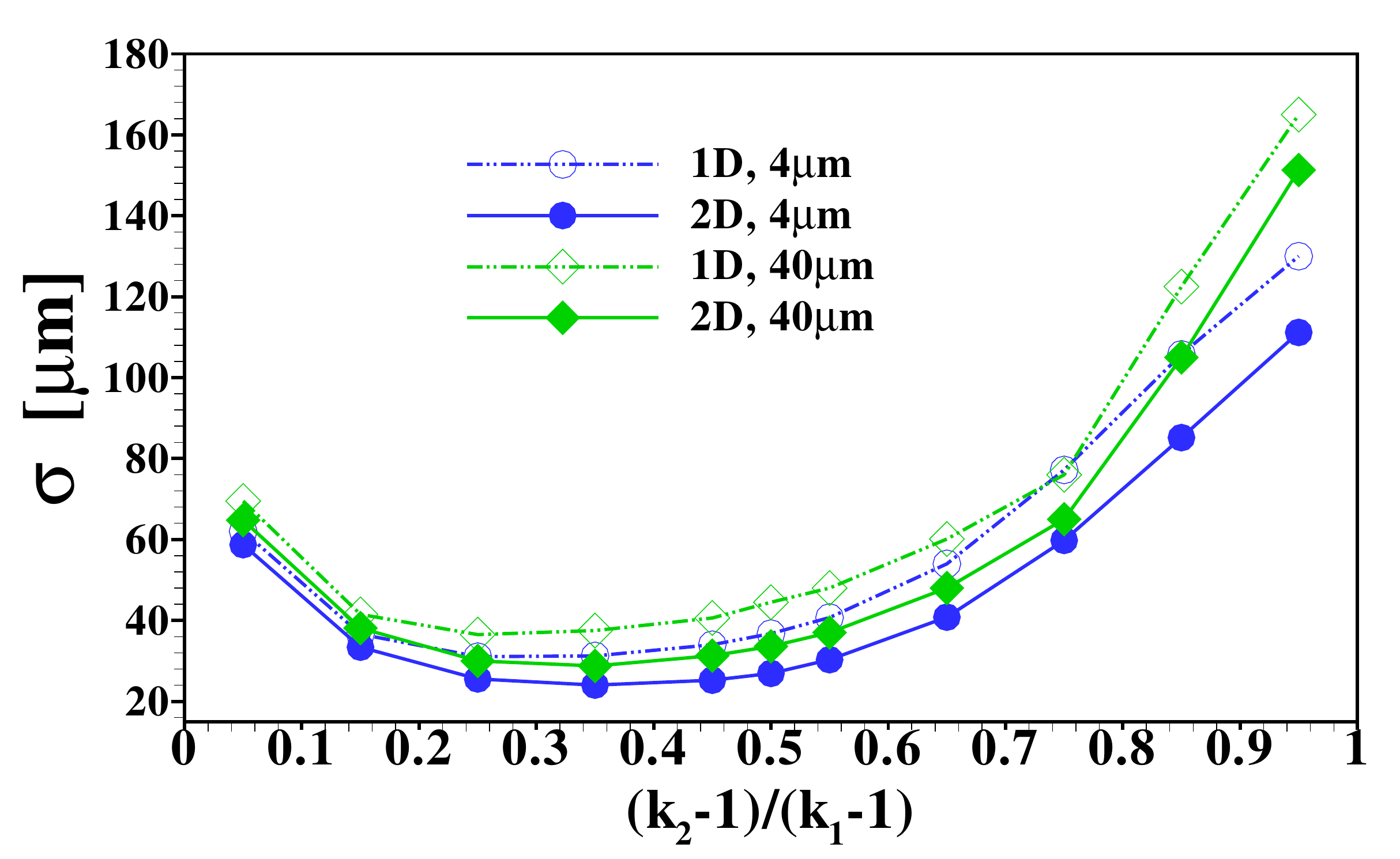}}\hspace{.25cm}
\subfigure[]{\label{c62}\includegraphics[height=3in]{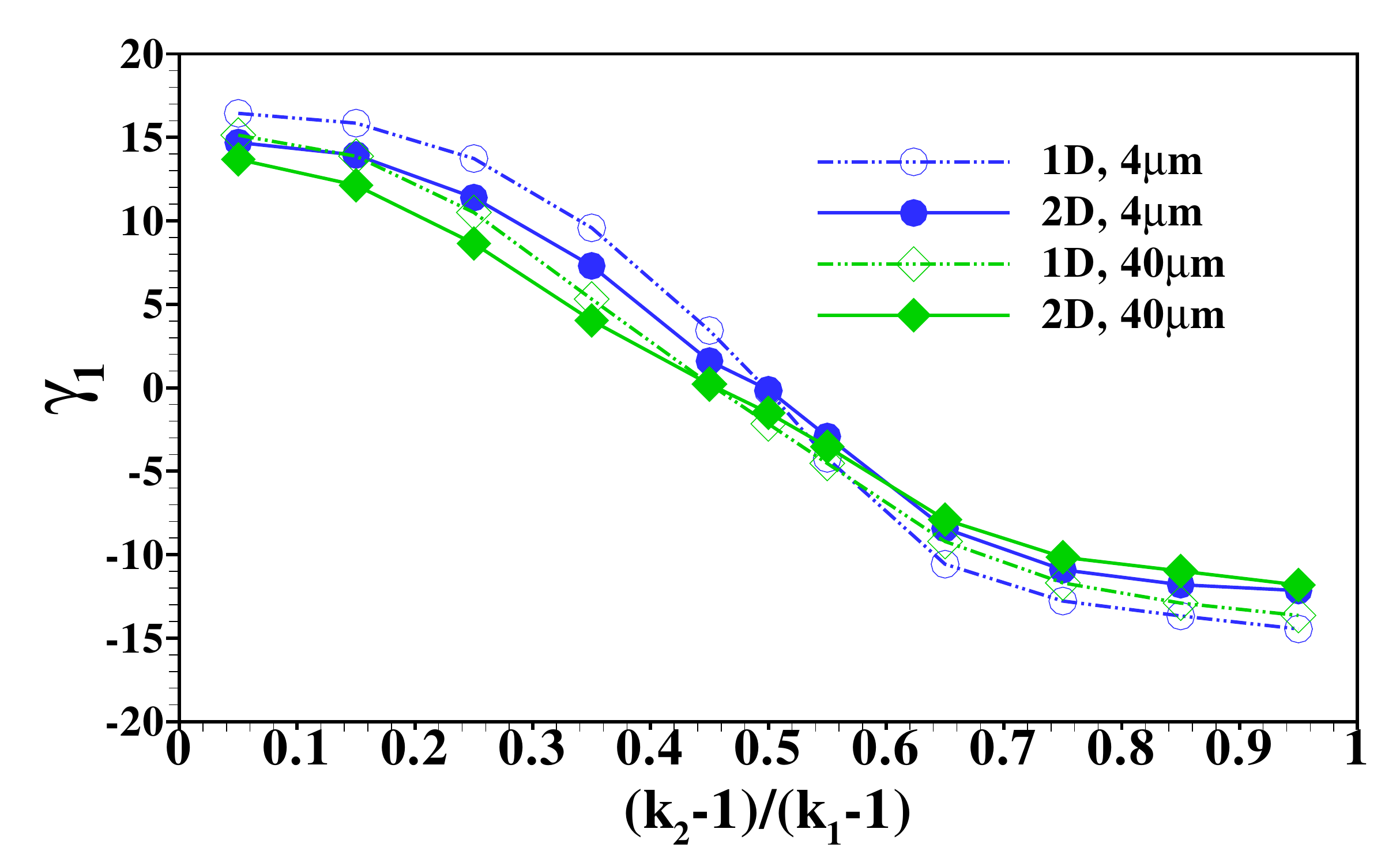}}\hspace{.25cm}
\caption{\label{fig:StDev}  Variation of the (a)  standard deviation and (b) skewness  of the sample distribution with $k_2(=\mu_{l}/\mu_s=D_{l}/D_s)$ for a fixed value of $k_1(=\mu_{l}/\mu_{t}=D_{l}/D_{t})=3$.  The P\'{e}clet number is $\mathrm{Pe}=50k_2$, the depth of the channel $H=25\,\mathrm{\mu m}$. The amount of sample is ${\cal{C}}_s/C_{l}^\infty=40\,\mathrm{\mu m}$ and $4\,\mathrm{\mu m}$. Lines with unfilled symbols represent the results from the 1D model (Taylor-Aris model), lines with filled symbols those from the 2D model.}
\end{figure}

\begin{figure}
\subfigure[]{\label{c91}\includegraphics[height=2.5 in]{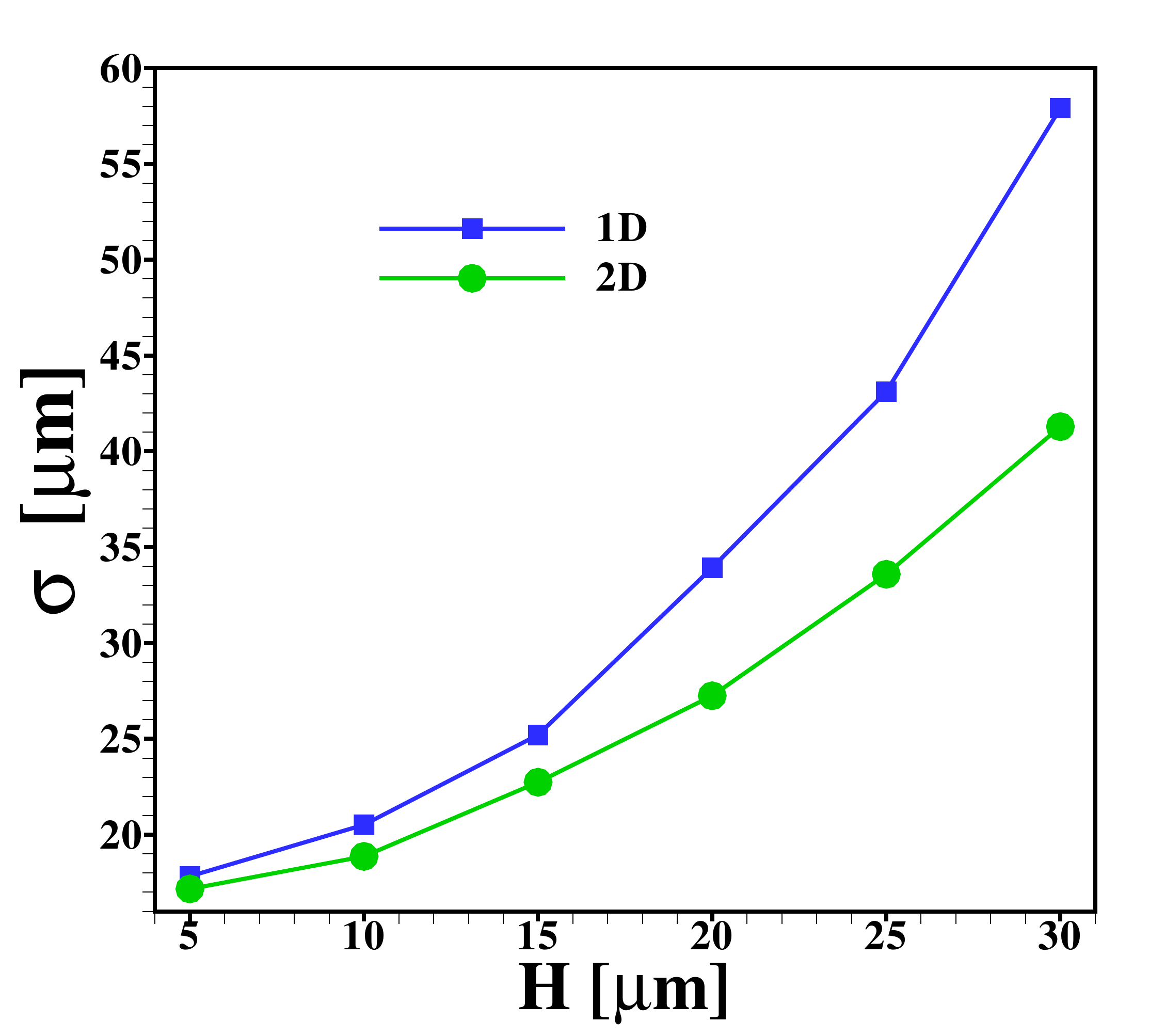}}\hspace{.25cm}
\subfigure[]{\label{c92}\includegraphics[height=2.5in]{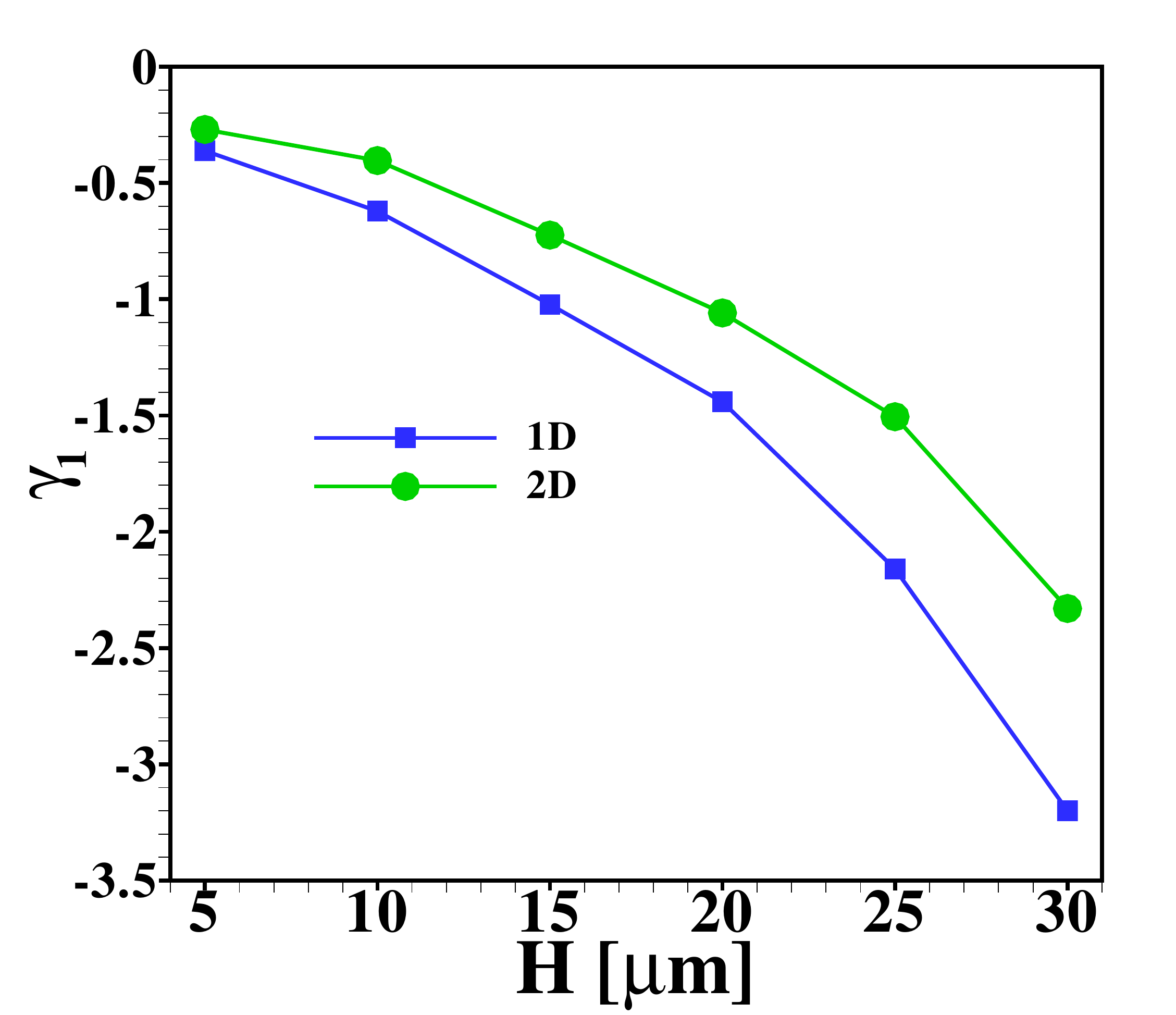}}\hspace{.25cm}
\caption{ \label{fig:StDev_H} Variation of the standard deviation of the sample distribution with the channel depth, where $k_1(=\mu_{l}/\mu_{t}=D_{l}/D_{t})=3$, $k_2(=\mu_{l}/\mu_s=D_{l}/D_s)=2$ and ${\cal C}_s/C^\infty_{l} = 40\,\mathrm{\mu m}$. The P\'{e}clet number varies with the channel depth as $\mathrm{Pe}=100\cdot H/(25\,\mathrm{\mu m})$. }
\end{figure}

The stage is now set for a more systematic study of the influence of the individual parameters on the form of the sample distribution. We use the width of the sample zone, $\sigma$, and its skewness, $\gamma_1$, in figures \ref{fig:StDev}, \ref{fig:StDev_allk} and \ref{fig:StDev_H} to investigate the dependence of the dispersion on diffusivity ratios of the individual electrolytes, the amount of sample present and the channel depth, respectively. 

In particular, figure \ref{fig:StDev_allk} shows the distribution's width and skewness for several values of $k_1$ when $k_2$ is varied in the range $1<k_2<k_1$. In order to compare the results we have plotted them using the rescaled variable $(k_2-1)/(k_1-1)$ so that the curves span the same range in parameter space. In this figure the remark already made in the discussion of figure \ref{fig:Variable_k2} is quantified: the sample dispersion is largest in the cases where the sample mobility is close to the mobility of one of the other electrolytes. This is reflected both in the increase of width, $\sigma$, as well as its skewness, $\gamma_1$, for values of $k_2$ close to either of the extreme cases 1 or $k_1$. The sign of the skewness reflects the fact that for $k_2$ close to 1, i.e. the diffusivity is very close to the LE diffusivity, the sample distribution has a long tail reaching into the LE, and similarly when the diffusivity is close to the TE diffusivity, i.e. $k_2$ close to $k_1$, the tail reaches to the left into the LE. We find the skewness to be almost zero, i.e. a symmetric sample distribution, when its mobility, $\mu_s$, is close to the harmonic average of $\mu_t$ and $\mu_l$. We note that the individual curves corresponding to different values of $k_1$ almost coincide showing that the dependence on $k_1$ is relatively weak when considering the rescaled $k_2$-values.

To further investigate the universality of the distribution width and skewness curves, we show in figure \ref{fig:StDev} the corresponding curves for different amounts of sample present in the system. In particular, we show results for a sample amount of ${\cal C}_s/C^\infty_{l} = 40\,\mathrm{\mu m}$ and ${\cal C}_s/C^\infty_{l} = 4\,\mathrm{\mu m}$ and find that both the width as well as the skewness of the distributions do not significantly change. As noted earlier, these two situations corresponds to plateau and peak mode in a situation of ITP without additional convective dispersion. This points towards the fact that the form of the distribution in dispersed plateau mode is not strongly dependent on the amount of sample present, the difference being mainly a scale factor proportional to the total amount, as already hinted at in the distributions shown in figure \ref{fig:Variable_k2}.

Both in figure \ref{fig:StDev_allk} and \ref{fig:StDev} we show data obtained within the 2D calculation as well as for the effective 1D Taylor-Aris dispersion model. As can be seen, there is good agreement between the two calculations corroborating the use of the simplified model for predicting area-averaged sample distributions in cases with convective dispersion. Nevertheless, it is important to keep in mind that the actual sample distribution within the channel will usually be richer in features in the 2D case, in particular we mention the focusing of the sample in the channel center observed for example in figures \ref{fig:VariableH_2D} and \ref{fig:Variable_ms_2D}.

%%%%%%%%%%%%%%%%%%%%%%%%%%%%%%%%%%

Another interesting aspect is how the width of the sample zone changes when varying the channel depth $H$, and hence the P\'{e}clet number, as displayed in figure \ref{fig:StDev_H}. From the effective diffusivity, eq.~(\ref{eq:EffDiff}), one expects a sample zone width that scales as $\sim \mathrm{Pe}^2$ in dispersed plateau mode. For large enough $H$, such that the dispersed plateau mode is indeed observed, this is reproduced well by the 1D model, as it should be. However, in the region shown this limit is not fully reached yet and only for values of $H$ larger than $30\; \mathrm{\mu m}$ the points follow $\sigma \sim H^2$ for the 1D results. As can be seen from the figure, the 2D model predicts less sample dispersion, a fact that can again be attributed to the focusing of sample in the channel center. Figure~\ref{fig:StDev_H}~(b) shows that the skewness of the distribution is correspondingly amplified by the increasing channel height.

\section{Conclusions}

We have investigated sample dispersion in situations in which electromigration in isotachophoresis is balanced by a Poiseuille counterflow. The focus was put on small sample amounts, i.e. situations when the area averaged sample distribution shows a peak. It is found that at large P\'{e}clet numbers, the sample dispersion is significantly increased compared to a situation without counterflow. Since the amount of sample is such that without the counterflow present one would typically obtain a short sample zone in plateau mode, we have termed this regime ``dispersed plateau mode''. One of the findings of this study is summarized in figures \ref{fig:StDev_allk} and \ref{fig:StDev} which show the dispersion of the sample zone as a function of the sample ion mobility. In particular, the diagram shows in which window the mobility of the sample can be chosen relative to the mobility of the surrounding electrolytes before significant broadening of the sample zone is observed. In particular, we find that the area averaged sample distributions are only weakly dependent on $k_1=\mu_l/\mu_t$ but vary strongly with $k_2=\mu_l/\mu_s$ (or more precisely with the rescaled value $(k_2-1)/(k_1-1)$). These results are also observed to be relatively insensitive to the amount of sample present in the system, cf. figure \ref{fig:StDev}, with the general form of the distribution remaining the same but the peak--height scaling with the total amount. These curves can thus be used to asses the degree of dispersion and skewness expected in an experiment. Similar results hold for the case in which it is desired to focus two sample zones stacked behind each other without the risk of too strong intermixing. 

Another key result of this study is that a 1D area-averaged model for the sample distribution based on Taylor-Aris dispersion agrees well with the more detailed 2D model in many situations. In view of the original derivation of the effective diffusion coefficient, the quality of the agreement is surprising, since the complex interplay of convection, electromigration and diffusion that shapes a transition zone in ITP is not accounted for. Since the computational cost of the 1D model is much less than that of the 2D model, this opens the door for fast simulations of ITP processes that are superimposed by convection. This could be of considerable relevance for a number of processes involving sample preconcentration and separation. However, it is important to stress that the detailed spatial distribution of the sample ions generally is more complex than the simplified 1D picture suggests. In particular we observe in the 2D calculations that the sample has a tendency to become concentrated in the channel center. This will have an impact in situations where different ionic species are made to react within an ITP experiment, as for example described in Goet et al.\cite{Goet_2009} and Bercovici et al.\cite{Bercovici_2012}, when long reaction times are attained via a Poiseulle counterflow.

\begin{acknowledgments}
T. Baier and S. Hardt kindly acknowledge support by the German Research Foundation (DFG) through the Cluster of Excellence 259. S. Hardt acknowledges helpful discussions with Juan Santiago, Stanford University.
\end{acknowledgments}

%%%%%%%%%%%%%%%%%%%%%%%%%%%%%%%%% Supplementary material %%%%%%%%%%%%%%%%%%%%%%%%%%%%%%%%%
\setcounter{figure}{0}
\makeatletter 
\renewcommand{\thefigure}{S\@arabic\c@figure}

%\pagebreak
\clearpage

%\newpage
\appendix
\section*{\Large Supplementary material}

\section*{S1 Discretisation of governing equations}
\begin{figure}% [thbp]
\begin{center}
\subfigure{\includegraphics[height=4.0in]{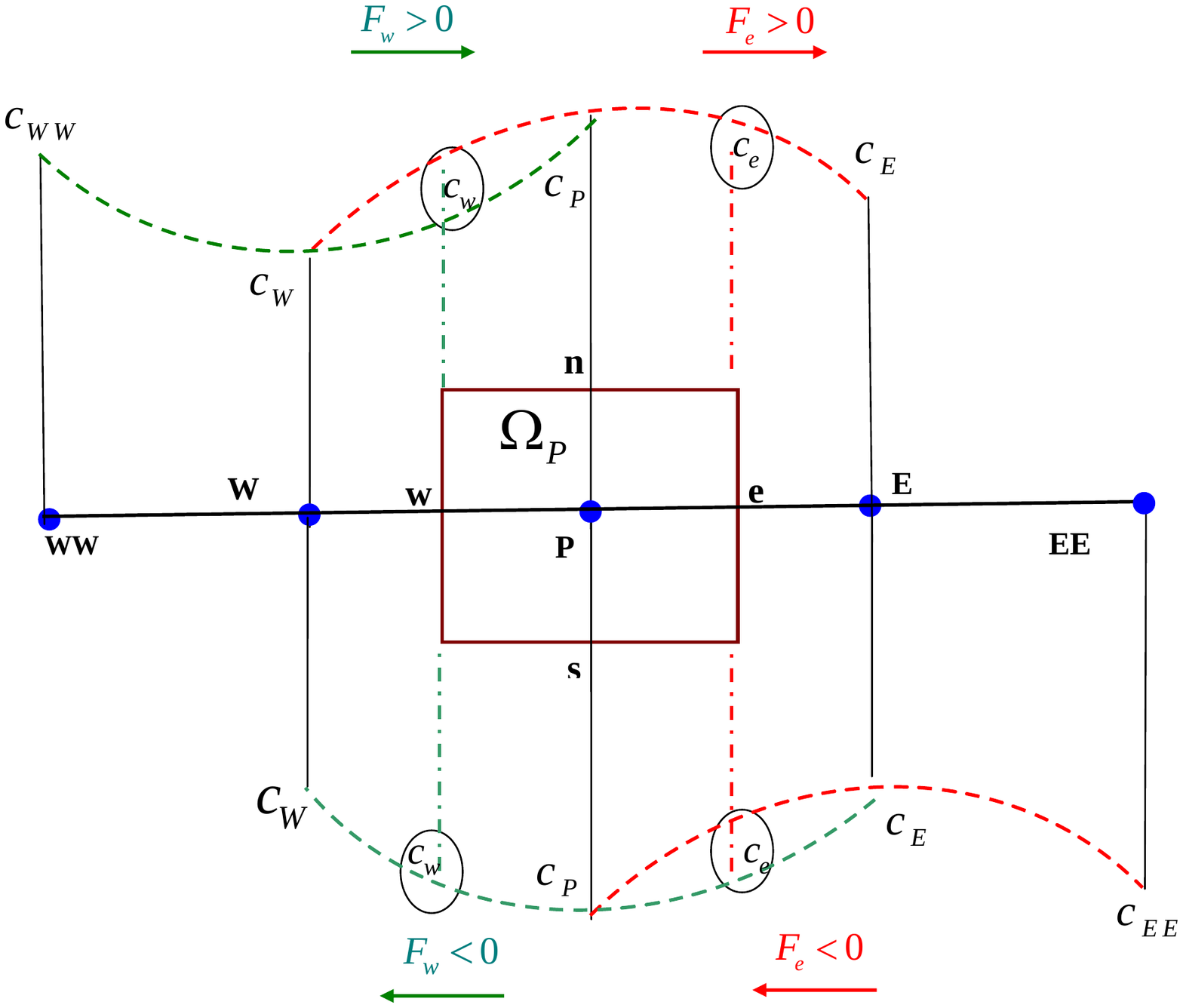}}
\caption{\label{fig:1} Schematic diagram for  control volume $\Omega_P$ and  interpolation for a variable $c$ based on the QUICK scheme. Here, $e$, $w$, $n$ and $s$ are the cell faces of the cell centered at $P$.}
\end{center}
\end{figure}

The Nernst-Planck  equation can be written as
\begin{equation}
\frac{\partial c}{\partial t} +\big[\frac{\partial}{\partial x}(Fc)+\frac{\partial}{\partial y}(Gc)\big]- \nabla^2 c=0
\end{equation}
where $F=u(y)- z_i\frac{D_{i}}{D_s}\frac{1}{Pe} \frac{\partial \phi}{\partial x}$, $G=- z_i\frac{D_{i}}{D_s}\frac{1}{Pe}\frac{\partial \phi}{\partial y}$ and $c$ is the concentration of the $i^{th}$ ionic species. The computational domain is sub--divided into a number of elementary rectangular cells  $\Omega_P$ with area $d\Omega_P$ whose sides are $dx_p$ and $dy_p$. The ion transport equations, when integrated over a cell $\Omega_P$ (cf. figure~\ref{fig:1}), yields the discretised form to advance the solution from $k^{th}$ time step to $(k+1)^{th}$ time step
$$
\frac{ c_P^{k+1}-c_P^{k}}{dt}~d{\Omega_P}+(F_e c_e-F_wc_w)\big|^{k+1}dy_P+(G_n c_n-G_sc_s)\big|^{k+1}dx_P$$
\begin{equation}
-\left[\frac{\partial c}{\partial x}\Big|_e-\frac{\partial c}{\partial x}\Big|_w\right]^{k+1}dy_P-\left[\frac{\partial c}{\partial y}\Big|_n-\frac{\partial c}{\partial y}\Big|_s\right]^{k+1}dx_P=0
\end{equation}
Here $n, s, e$ and $w$ refer to the northern, southern, eastern, western face of the cell (cf. figure~\ref{fig:1}). An implicit first--order scheme is used to discretise the time derivatives present. The electromigration and convection terms are discretised through a higher order upwind, QUICK scheme as follows

\begin{equation}
\begin{aligned}
(F_e c_e-F_wc_w)dy_P &+ (G_n c_n-G_sc_s)dx_P=\\
&\Big[ (\frac{3}{8}c_E+\frac{3}{4}c_P-\frac{1}{8}c_W)[[F_e,0]] -  (\frac{3}{4}c_E+\frac{3}{8}c_P-\frac{1}{8}c_{EE}) [[-F_e,0]] \Big]dy_P \\
-&\Big[ (\frac{3}{8}c_P+\frac{3}{4}c_W-\frac{1}{8}c_{WW})[[F_w,0]]- (\frac{3}{4}c_P+\frac{3}{8}c_W-\frac{1}{8}c_{E}) [[-F_w,0]]           \Big]dy_P \\
+&\Big[ (\frac{3}{8}c_N+\frac{3}{4}c_P-\frac{1}{8}c_S)[[G_n,0]]-  (\frac{3}{4}c_N+\frac{3}{8}c_P-\frac{1}{8}c_{NN}) [[-G_n,0]]           \Big]dx_P \\
-&\Big[ (\frac{3}{8}c_P+\frac{3}{4}c_S-\frac{1}{8}c_{SS})[[G_s,0]]-  (\frac{3}{4}c_P+\frac{3}{8}c_S-\frac{1}{8}c_{N}) [[-G_s,0]]           \Big]dx_P
\end{aligned}
\end{equation}
where the operator $[[a,b]]$ yields the larger of $a$ and $b$. The diffusion flux  at interfaces `$e$' and '$w$' are evaluated as
\begin{equation}
\frac{\partial c}{\partial x}\Big|_e=\frac{c_{E}-c_{P}}{0.5(dx_P+dx_E)}
\end{equation}
and
\begin{equation}
\frac{\partial c}{\partial x}\Big|_w=\frac{c_{P}-c_{W}}{0.5(dx_P+dx_W)}.
\end{equation}
A similar procedure is adopted for estimating the variables at the other cell faces `$n$' and `$s$'.  Note that the big letter subscripts denote the cell centers in which variables are stored and small letter subscripts denotes the corresponding cell faces. Since the ion transport equations are coupled with the charge conservation equation, we adopt an iterative method. In order to reduce the system of algebraic equations into a tri--diagonal form, the variables at the locations `$EE$'  and `$WW$' are taken as the previous iterated values. At every time--level, the solutions are obtained iteratively.

At every time--level the iteration procedure starts with a guess for the electric potential at each cell center. At every iteration, the charge conservation equation is integrated over each control volume $\Omega_P$ through the finite volume method. The elliptic PDE for charge conservation is solved by a line--by--line iterative method along with the successive--over--relaxation (SOR) technique. The iterations are continued until the absolute difference between two successive iterations becomes smaller than the tolerance limit $10^{-6}$ for concentration as well as for the electric potential.

\clearpage
\section*{S2 Boundary Conditions}

\begin{figure}[th!]
\begin{center}
\subfigure{\includegraphics[height=2 in]{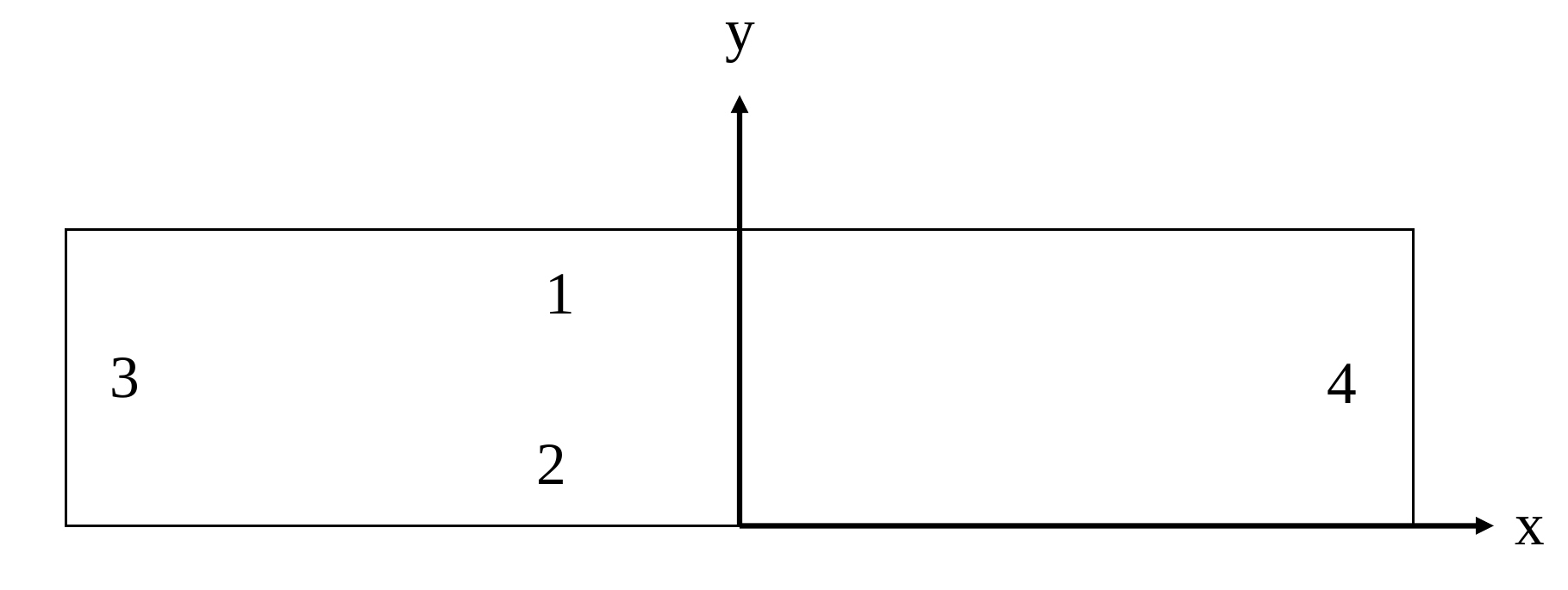}}
\caption{\label{fig:BC} Schematic of the computational domain.}
\end{center}
\end{figure}
The channel walls (boundaries 1 and 2 in figure~\ref{fig:BC}) are impermeable for all ions
\begin{equation}
\textbf{N}_i.\textbf{n}=0,~~\nabla \phi.\textbf{n}=0
\end{equation}
Here the subscript $i=l$, $s$, $t$ refers to LE, sample and TE and $\textbf{n}$ is the unit outward normal on the wall.

Along the inlet (boundary 3) the concentrations and electric potential are prescribed as
\begin{equation}
c_{t}=\frac{\mu_{l}+\mu_0}{\mu_{t}+\mu_0}\frac{\mu_{t}}{\mu_{l}},~~c_{l}=c_{s}=0,~~ \phi=-\frac{E_tH}{\phi_0}x_3
\end{equation}
while along the outlet (boundary 4) we set
\begin{equation}
c_{t}=c_{s}=0,~~ c_{t}=1,~~\phi=-\frac{E_lH}{\phi_0}x_4,
\end{equation}
where $x_3=-x_4$ are the positions of the boundaries, located symmetrically around the origin.

%\clearpage
\section*{S3 Effect of grid size, time evolution and validation}

\begin{figure} %[thbp]
\begin{center}
\subfigure{\includegraphics[height=3in]{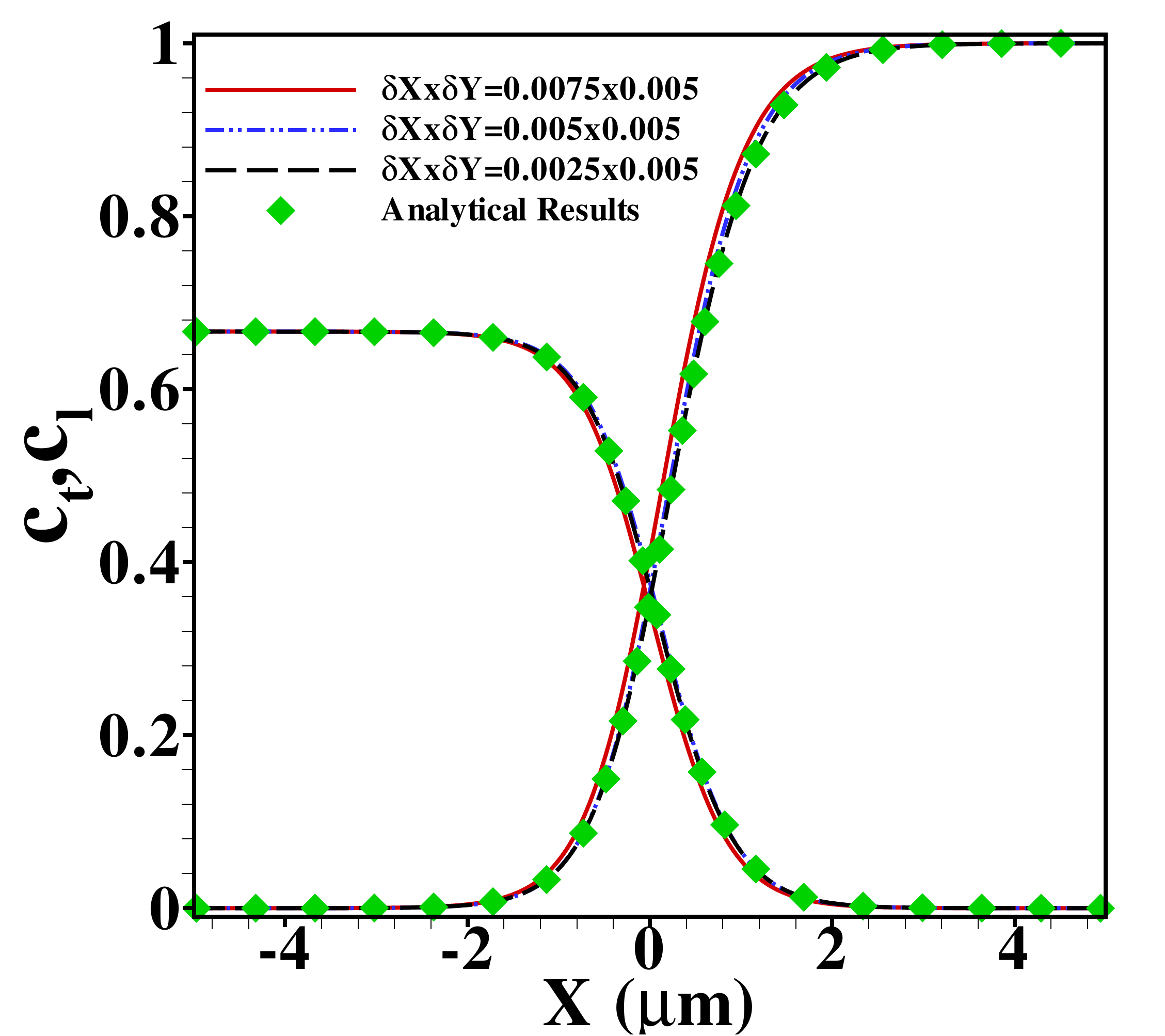}}
\caption{\label{fig:grid} Effect of grid size on the TE and LE concentration profiles and comparison with the analytical solution due to Goet et al. \cite{14}. The electrophoretic mobility ratio is taken as $k_1(=\mu_l/\mu_t=D_l/D_t)=2$.}
\end{center}
\end{figure}

For code validation we consider pure ITP without a sample zone, i.e. only LE and TE present, since for this case an analytic result is available. This analytical solution due to Goet et al. \cite{14} for the TE concentration and axial electric field in a co--moving frame of reference under no bulk fluid flow can be expressed through the hypergeometric function as
\begin{equation}\label{eq:ct_ct}
\frac{C_t(x)}{C_t^\infty}=F\left[1,\frac{E_t}{\Delta E}; 1+\frac{E_t}{\Delta E}; -\frac{\mu_l+\mu_0}{\mu_t+\mu_0}e^{\frac{\Delta E}{\phi_0}x}\right]
\end{equation}
and
\begin{equation}
\frac{E_t}{E(x)}=\frac{C_{t}(x)}{C^{\infty}_{t}}\left[1+\frac{\mu_l+\mu_0}{\mu_t + \mu_0}e^{\frac{\Delta E}{\phi_0}}\right]
\end{equation}
At steady state the concentrations of TE and LE are related via
\begin{equation}\label{eq:clct}
\ln\left(\frac{C_{l}}{C_{t}}\right)=\frac{\Delta E}{\phi_0} x
\end{equation}
where $\Delta E=E_t-E_l$. Using relation (\ref{eq:ct_ct}), the LE concentration can easily be obtained from (\ref{eq:clct}).

Figure \ref{fig:grid} shows the effect of grid size on the TE and LE concentration profiles for a pure ITP without sample obtained with our code, together with the analytical results described above. As in the main text, the results are for $D_l=7.0\times10^{-10}m^2/s$, $E_0=10^5V/m$, $l_t=l_l=1/2$ and a bulk LE concentration of $C^\infty_{l}=10^{-3}\,\mathrm{M}$. The mobility ratio is taken to be $k_1=\mu_l/\mu_t=2$. Results are shown for three different grid sizes, with increasing fineness of grid in the $X$-direction. As can be seen, the analytical and numerical results agree very well for a grid of spacing $\delta X=\delta Y = 0.005$.

\begin{figure}% [thbp]
\begin{center}
\subfigure{\includegraphics[height=2.6 in]{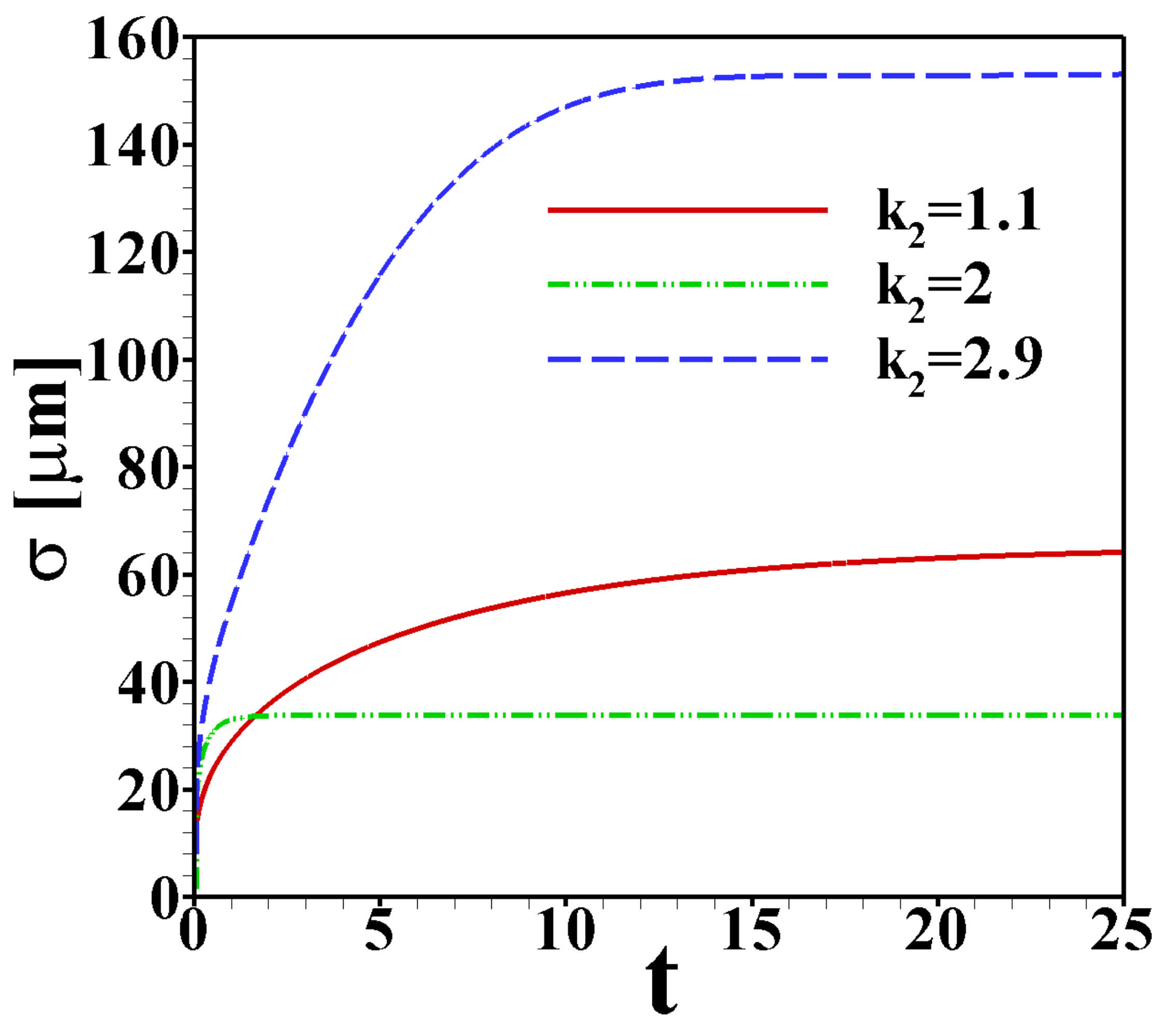}}
\caption{\label{fig:time} Time evolution of the standard deviation, $\sigma$, reflecting the sample dispersion for ITP with Poiseuille counter--flow. Here $k_1(=\mu_l/\mu_t=D_l/D_t)=3$ with $k_2(=\mu_l/\mu_s=D_l/D_s)=1.1$, $2$ and $2.9$ and the amount of sample is ${\cal C}_s/C_l^\infty=40\,\mu m$.}
\end{center}
\end{figure}

Figure \ref{fig:time} shows the time evolution towards the steady state for the case of ITP with a sample zone held stationary in the middle of the channel by a Poiseuille counterflow. As indicator the standard deviation, $\sigma$, describing the width of the sample zone, is used here. The three cases shown correspond to values of the mobility ration $k_2=\mu_l/\mu_s$ close to the edges and in the middle of its range of validity, $1<k_2<k_1$. In all simulations shown in the main text care was taken that the steady state has indeed been reached.

\section*{S4 TE, sample and LE concentration profiles}

\begin{figure}% [thbp]
\begin{center}
\subfigure[]{\includegraphics[height=1.8 in]{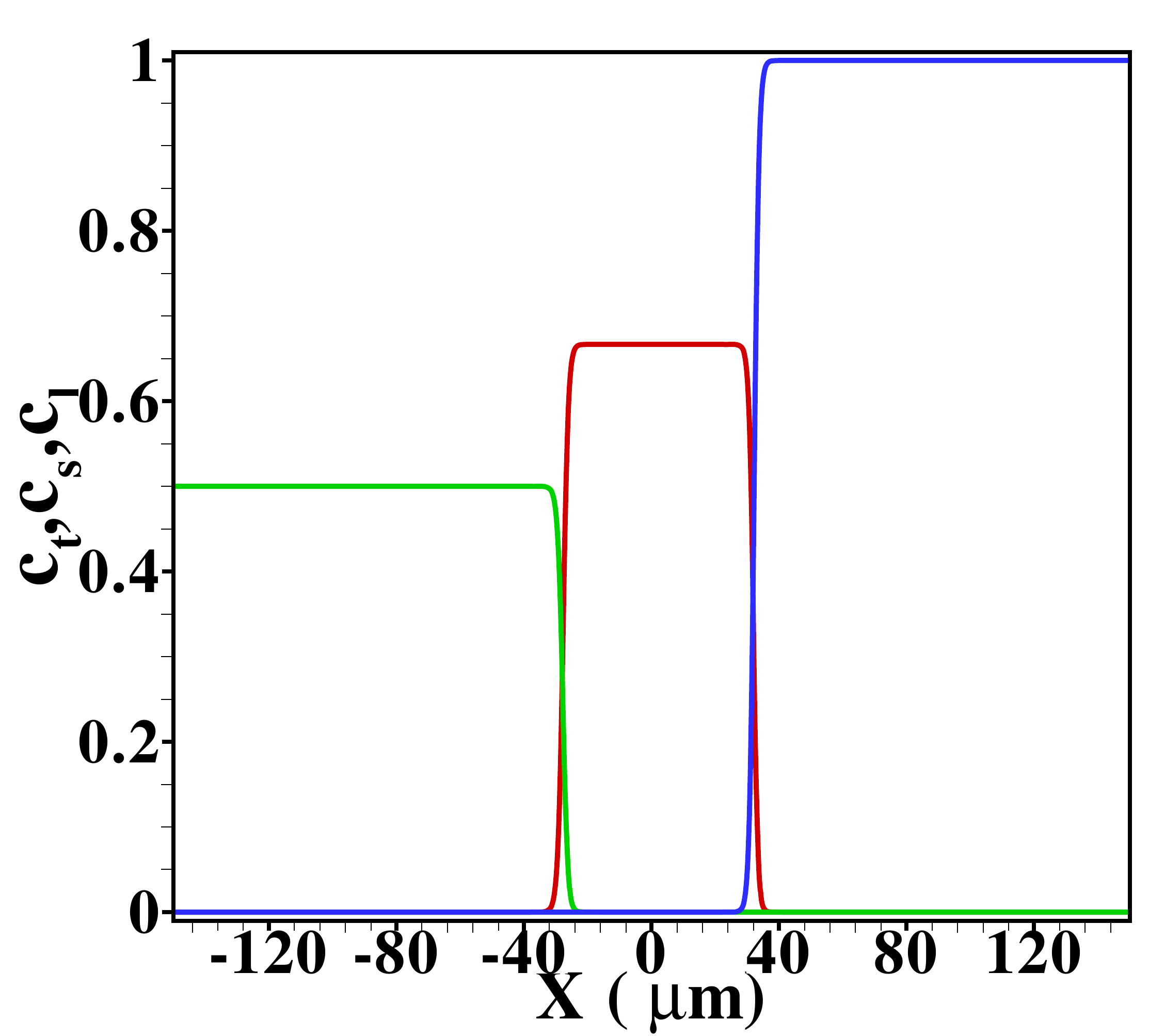}}
\subfigure[]{\includegraphics[height=1.8 in]{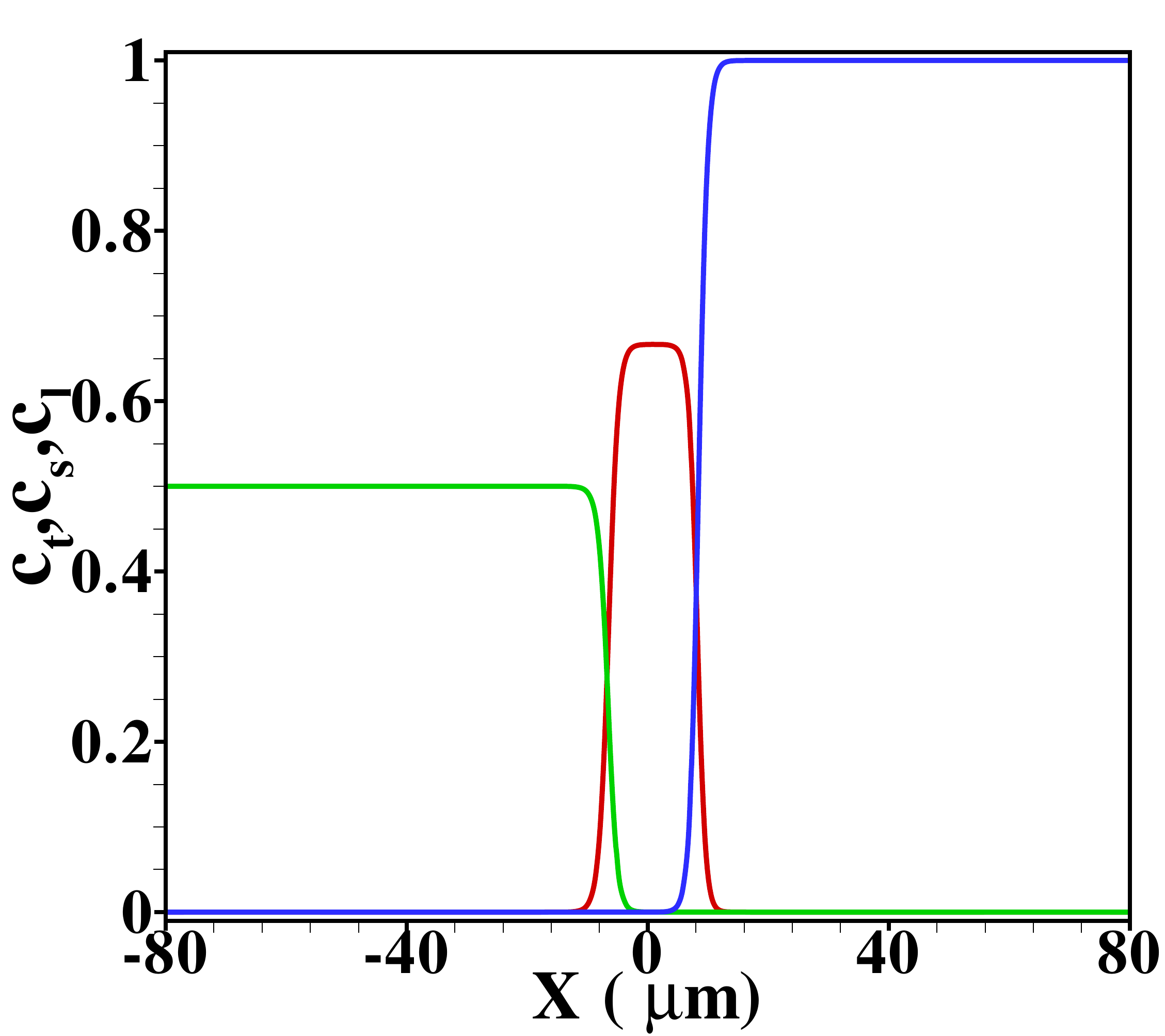}}
\subfigure[]{\includegraphics[height=1.8 in]{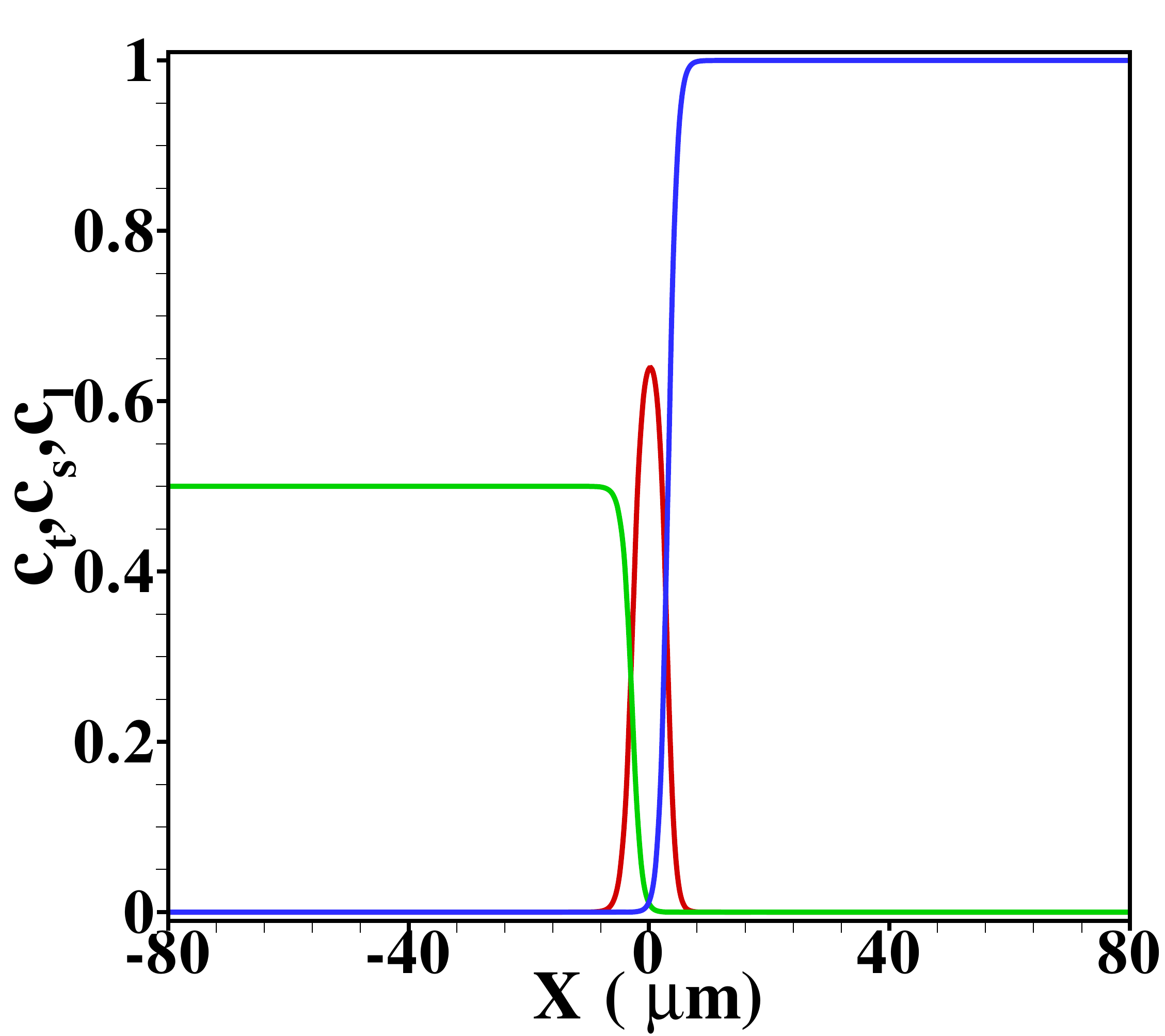}}
\caption{\label{fig:pureITP_varSample} Variation of TE (green), sample (red) and LE (blue) concentrations at steady state in a frame of reference moving at  the constant speed $U^{ITP}$ for three different values of sample  amount (a) ${\cal C}_s/ C_l^\infty=40\,\mu m$; (b) ${\cal C}_s/C_l^\infty=10\,\mu m$ and (c) ${\cal C}_s/C_l^\infty=4\,\mu m$. Here the mobility ratios are taken as  $k_1(=\mu_l/\mu_t=D_l/D_t)=3$  and $k_2(=\mu_l/\mu_s=D_l/D_s)=2$.}
\end{center}
\end{figure}

For reference, we show in figure \ref{fig:pureITP_varSample} the TE, sample and LE concentration profiles obtained for $l_l=l_t\approx 0.5$ and $\l_s \approx 0$ in the case without Poiseuille counterflow. The cases shown range from sample amounts of ${\cal C}_s/C_l^\infty=40\,\mu m$ to ${\cal C}_s/C_l^\infty=4\,\mu m$, two typical cases shown in the main text. In the former case the sample concentration clearly develops a plateau, while for the latter a Gaussian peak is observed. Note that both cases are in the regime of dispersed plateau mode, i.e. the area averaged concentration distribution shows a (distorted) Gaussian peak when Poiseuille counterflow is applied to keep the sample zone stationary.

\begin{figure}% [thbp]
\begin{center}
\subfigure[]{\includegraphics[height=1.8 in]{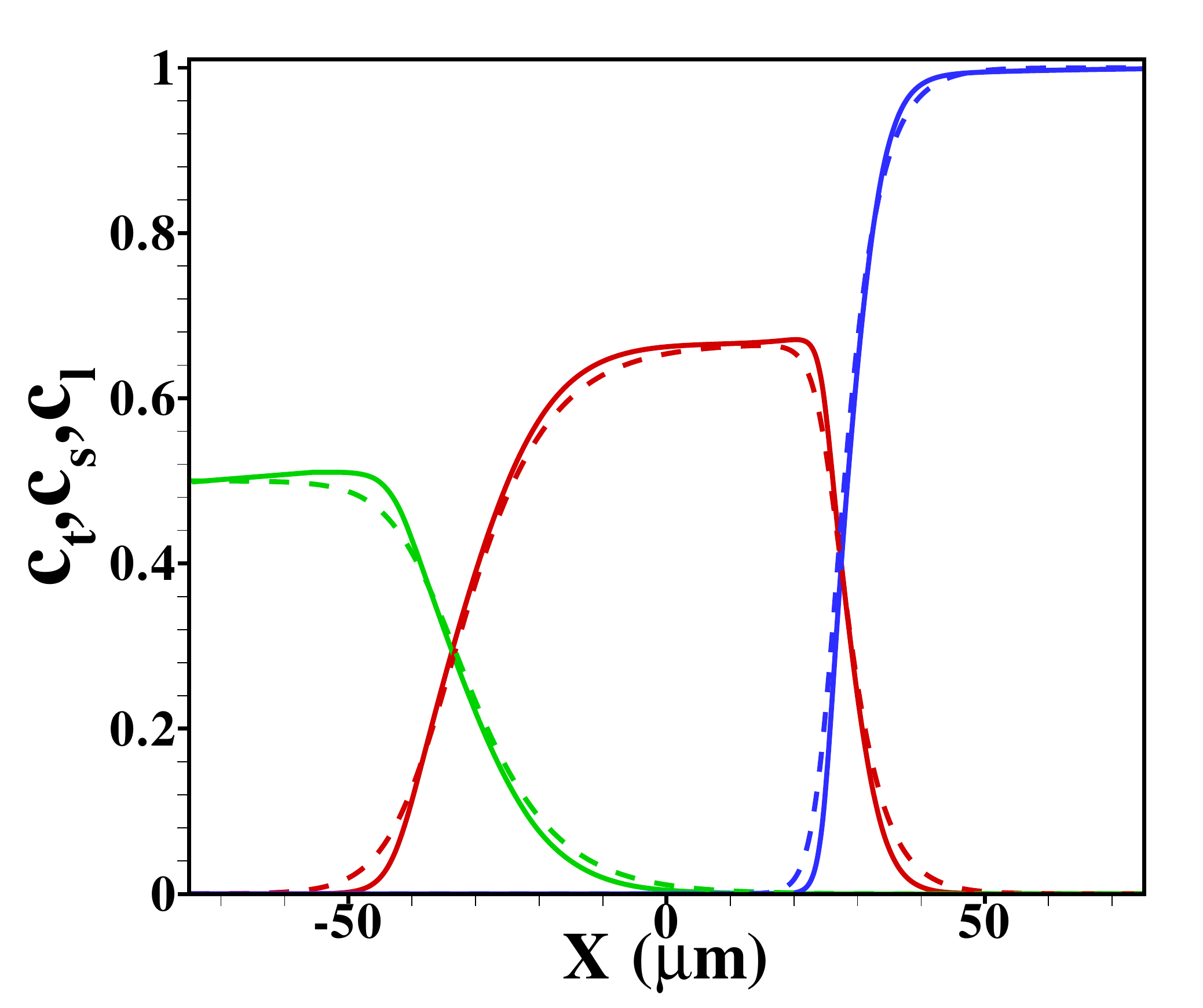}}
\subfigure[]{\includegraphics[height=1.8 in]{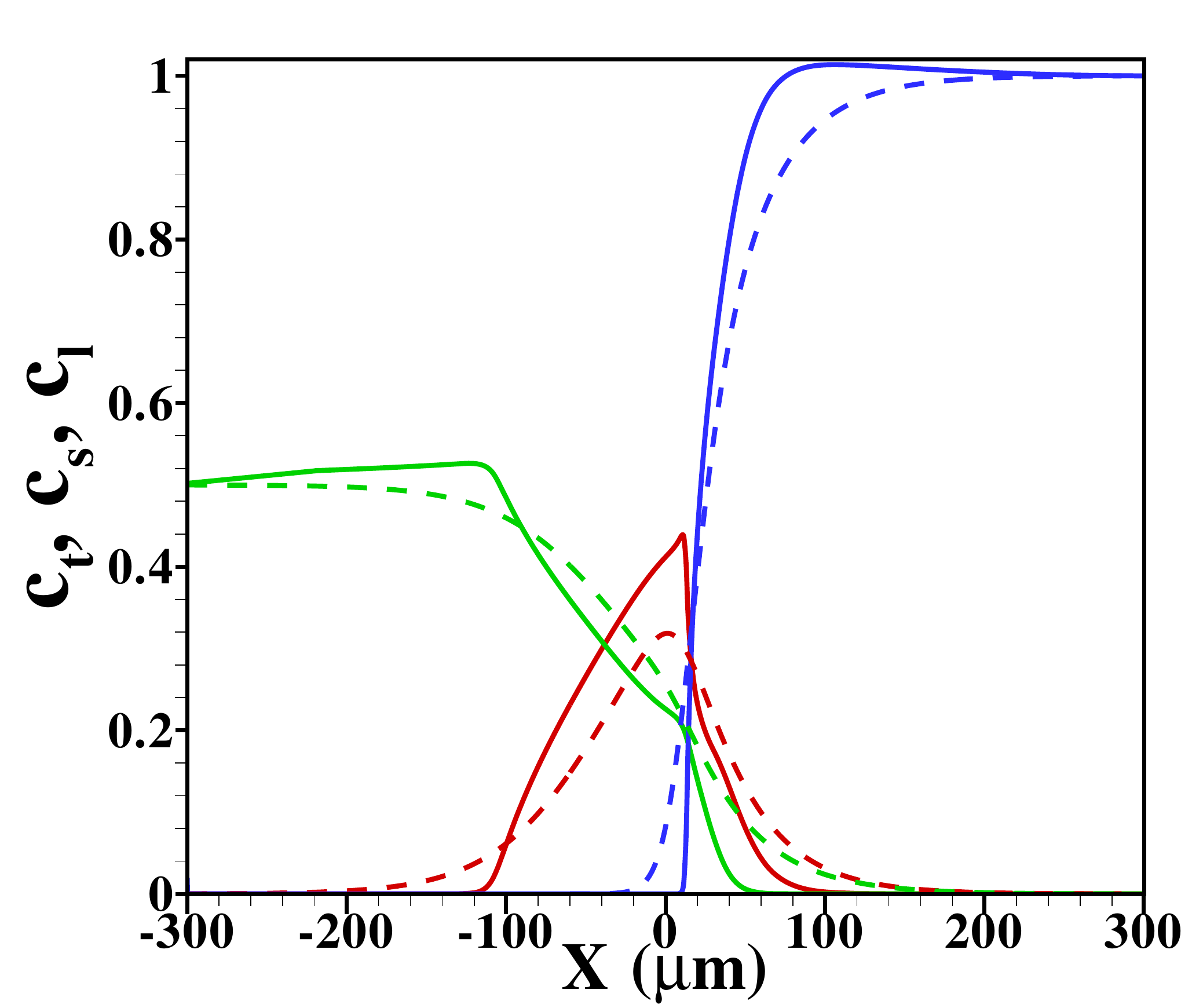}}
\caption{\label{fig:sITP_VariableH_1D} Area-averaged concentration distributions during ITP in Poiseuille counterflow of TE (green), sample (red) and LE (blue) for $k_1(=\mu_l/\mu_t=D_l/D_t)=3$  and $k_2(=\mu_l/\mu_s=D_l/D_s)=2$ with channel depths of (a)~$H=10\,\mathrm{\mu m}$ and (b) $H=30\,\mathrm{\mu m}$. Full and dashed lines correspond to results obtained with the 2D and 1D models, respectively. The amount of sample is ${\cal{C}}_s/C_{l}^\infty=40\,\mathrm{\mu m}$. (Cf. figure 3 in the main text, where only the sample distribution is shown.)}
\end{center}
\end{figure}

In the main text the primarily focus is on the distribution of the sample ions in the channel. An example for the effect on the LE and TE concentrations in the situation with Poiseuille counterflow is shown in figure \ref{fig:sITP_VariableH_1D}. The area-averaged concentrations of all three ions (TE, sample and LE) is shown in the case of mobility ratios $k_1=\mu_l/\mu_t=3$, $k_2=\mu_l/\mu_s=2$ and sample amount of ${\cal{C}}_s/C_{l}^\infty=40\,\mathrm{\mu m}$ for channel depths of $H=10\,\mathrm{\mu m}$ and $30\,\mathrm{\mu m}$, respectively. As such the situation corresponds exactly to the one shown in figure 3 of the main text. It is evident, that the 1D area-averaged model is able to capture the results obtained with the 2D model with a similar accuracy for the LE and TE as for the sample ions.

%\clearpage
%\begin{thebibliography}{99}
%\bibitem{Goet_2011} G. Goet, T. Baier, S. Hardt, ``Transport and separation of micron sized particles at isotachophoretic transition zones", Biomicrofluidics 5, 014109 (2011).
%\end{thebibliography}

\end{document}